\documentclass[twocolumn]{article}
\usepackage{graphics}
\usepackage{amsmath}    % need for subequations
\usepackage{amssymb}    % need for subequations
\usepackage{graphicx}   % need for figures
\usepackage{verbatim}   % useful for program listings
\usepackage{color}      % use if color is used in text
\usepackage{subfigure}  % use for side-by-side figures
\usepackage{hyperref} 
\usepackage{appendix} 
\newcommand{\mb}{\mathbf}
\newcommand{\bs}{\boldsymbol}

\begin{document}
\title{Self propulsion of droplets driven by an active permeating gel}
\author{R. Kree and  A.Zippelius\\ Georg-August-Universit\"at G\"ottingen, 
Institut f\"ur Theoretische Physik,\\Friedrich-Hund-Platz 1, 37077 G\"ottingen, Germany}% 
%{\today}

\maketitle

\begin{abstract}
We discuss the flow field and propulsion velocity of active
  droplets, which are driven by body forces residing on a rigid gel.
  The latter is modelled as a porous medium which gives rise to
  permeation forces. In the simplest model, the Brinkman equation,
  the porous medium is characterised by a single length scale $\ell$ --
  the square root of the permeability.  
  We compute the flow fields
  inside and outside of the droplet as well as the energy dissipation
  as a function of $\ell$. 
  We furthermore show that there are  optimal gel fractions, giving rise to maximal linear and rotational velocities.  
%   and show that the propulsion velocity of the
%  droplet is nonmonotonic as a function of this lenghtscale.   
  In the
  limit $\ell\to\infty$, corresponding to a very dilute gel, we
  recover Stokes flow. The opposite limit, $\ell\to 0$, corresponding
  to a space filling gel, is singular and not equivalent to Darcy's
  equation, which cannot account for self-propulsion
%
%      {47.63.Gd}{swimming microorganism}   \and
%      {47.56+r}{flow through porous media} \and
%      {87.17.Jj}{cell locomotion, chemotaxis} 
%     } % end of PACS co
\end{abstract} %end of abstract
\maketitle
\section{Introduction}
\label{intro}
Hydrodynamic models of self-propelled particles \cite{Lauga_2009} have focused primarily
on either squirmers built on ciliary propulsion or phoretic effects --
both localised on the surface of the swimmer. Here we consider instead
active forces in the interior of a living organism or cell. The
complex spatial organisation of a cell includes a network of
filaments, which serves not only as a backbone but also provides the
tracks for active transport by motor proteins. Biomolecules, vesicles
or even organelles can be actively transported in this way, thereby
generating flow in the cytosol. Cytoplasmic streaming has been
discussed extensively~\cite{Goldstein2015} in the context of intracellular
transport. However sustained flow of the cytosol can have various
other effects. Here we discuss the possibility of self-propulsion:
Active forces generate internal flow which in turn gives rise to
swimming motion \cite{Kree_2017}.

Coarse grained models of the cell interior are frequently based on
multi-component or multi-phase systems, in the simplest case a two
fluid model~\cite{Alt1999} or a biphasic system, representing the
filamentary network and the interstitial fluid~\cite{Spitzer2013}. We
built on this work and model the cytoplasm as a biphasic system,
consisting of a gel and a sol phase.  The gel is modeled as a porous
medium and the sol as a Newtonian fluid. Poroelastic materials as a
model for the cytoplasm~\cite{Moeendarbary2013} have been discussed
i.a. in the context of force-indentation curves of cells. Here, we
assume that the active forces are localised on the gel and drive
permeation flows in the sol~\cite{Juelicher2011}. As a first step, we
ignore the elasticity of the porous gel and assume the gel to be
rigid. Within this approximation the dynamics of the gel is restricted
to rigid body motion and the flow of the fluid component follows
Brinkman's equation. The resulting model can be solved analytically,
allowing us to compute the linear and angular velocity of propulsion
as well as the flow fields inside and outside of the droplet.

Our study gets further motivation from increasing experimental efforts
to construct simplified biomimetic devices or even synthesize
artificial cells with controlled ingredients and the ability for
self-assemblance, active processes and some degree of functionality.
Mixtures of actin filaments, cross-linkers and motor proteins exhibit
stationary dynamic states with high
activity~\cite{Koehler2011}. Active nematics of microtubules and
motors inside lipid vesicles also exhibit dynamic states, sometimes
periodoc and possibly shape changing~\cite{Keber2014}. More recently
droplet stabilised vesicles~\cite{Spatz2017} have been assembled and
loaded with biomolecules to synthesize increasingly more complex,
functional cells in aequous solution. Theoretical approaches have
suggested that active droplets may serve as models for
protocells~\cite{Zwicker2}. Droplets generated by phase separation
provide compartments for the spatial organization of chemicals with
their size being controlled by chemical reaction
rates\cite{Zwicker1}. Active droplets were furthermore shown to
spontaneously split into two identical daughter cells~\cite{Zwicker2},
thus behaving similarly to cells. Here we focus on the active motion
of biphasic droplets.

The paper is organized as follows. We introduce the model in
sec.~\ref{model}. Linear and rotational motion of the droplet as well
as the interior flow fields are presented in sec.~\ref{sec:linear} and
sec.~\ref{sec:rotation}. The singular limit of very small permeation
length is dicussed in sec.~\ref{sec:limits}, where we also show that
Darcy's equation cannot account for self-propulsion. Finally, in
sec.~\ref{sec:conclusions} we present our conclusions and give an
outlook to future work. Details of the calculation are delegated to 3
appendices.

\section{Model}
\label{model}
\begin{figure}
 \begin{center}
 \includegraphics[width=0.6\linewidth]{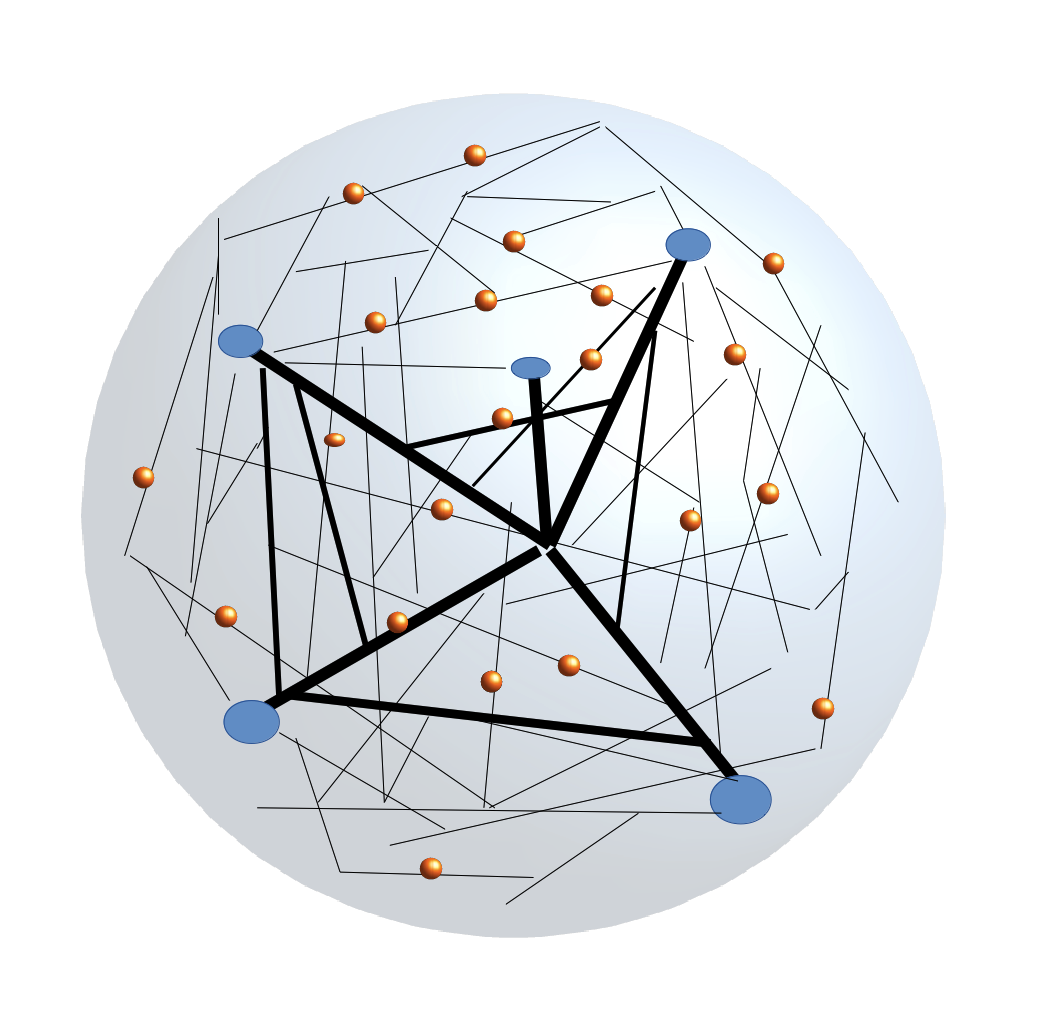}
 \end{center}
 \caption{\label{fig:model} Our model is a continuum version of the scetched situation, where actively transported particles (for example vesicles) move along the rigid filaments of a gel. }
 \end{figure}

We consider a neutrally buoyant spherical droplet of radius $R=1$, immersed into an
incompressible Newtonian fluid of viscosity $\eta^+$. The interior of
the droplet is composed of a statistically homogeneous and isotropic rigid gel  coexisting with another
incompressible fluid of viscosity $\eta$. Both the internal fluid and the gel are assumed to be completely immiscible in the ambient fluid. The volume fraction occupied by the gel (gel fraction) will be denoted by $\phi$.  The disorder or volume averaged flow velocity field $\mathbf{v}$ inside the droplet is assumed to obey the Brinkman
equation~\cite{Brady_1987,Rubinstein_1989}
\begin{equation}
	\nabla^2 \mathbf{v}(\mb{r})-\frac{1}{\kappa(\phi)}(\mb{v}(\mb{r})- \mb{v}^{gel})=\frac{1}{\eta}\Big(\nabla p(\mb{r})- \mb{f}^{act}(\mb{r},\phi)\Big).
	\label{eq:Brinkman1}
\end{equation} 
Here $\kappa(\phi)$ denotes the intrinsic permeability, which depends on the details of the gel structure, the pressure field $p(\mb{r})$ is determined from the incompressibility constraint $\mb{\nabla}\cdot\mb{v}=0$ and the body force density $\mb{f}^{act}$ is due to active processes in the gel, which drive permeation flows. As an example, one may think of cargo particles actively moving on a random network of stiff filaments, thereby generating flow (see sketch in Fig \ref{fig:model}).   We introduce a frame of reference with its origin at the center of the sphere. In this frame  the velocity of the rigid gel must be of the form 
\begin{equation}
\mb{v}^{gel}(\mb{r})=\mb{v}_g + \boldsymbol{\omega}_g\times\mb{r}
\label{eq:gelvelocity}
\end{equation}
with constant vectors $\mb{v}_g$ and $\boldsymbol{\omega}_g$. 

All passive properties of the gel, which influence the flow field
$\mb{v}$ are subsumed in the permeability $\kappa$ and the vicosity
$\eta$. In polymeric gels, the length scale $\sqrt{\kappa}=\ell$ is
also referred to as the hydrodynamic screening length. There are many
models in the literature \cite{Tamayol2011}, which relate $\kappa$ to
other macroscopic properties of the gel (for example gel fraction,
specific surface area and tortuosity).  Here, we restrict our
attention to the dependence of $\kappa$ on the gel fraction $\phi$.
Without any additional information about the gel structure, we
consider $\kappa(\phi)$ as a smooth monotonic function, which varies
between the obvious limiting cases (i) $\kappa\to 0$ for $\phi\to 1$
and (ii) $\kappa\to \infty$ if $\phi\to 0$. $\kappa$ may vanish before
$\phi=1$ is reached, if the fluid no longer percolates.
%gel structure possesses a percolation threshold
%$\phi_p<1$,
Here we are not interested in the behaviour close to the percolation
threshold and choose a variant of the classical Carman-Kozeny
formula~\cite{Carman_1956} 
\begin{equation}
\label{eq:kappa}
\kappa(\phi)=\ell^2=\ell_0^2(1-\phi)^2/\phi^2
\end{equation}
for purposes of illustration. Our qualitative results do not depend on
the precise form of $\kappa(\phi)$.
As the force density
$\mb{f}^{act}(\mb{r},\phi)$ is assumed to result from active elements, which are distributed smoothly over the gel, we assume that 
$\mb{f}^{act}(\mb{r},\phi)\propto\phi$.

 Outside of the volume $V$, the flow velocity field $\mathbf{v}^{+}$
 satisfies Stokes equation without the permeation force and without
 active forces. Boundary conditions at an interface $\partial V$
 between a porous medium and a Newtonian fluid are still discussed in
 the literature. Various different forms have been proposed
 \cite{Ochoa_1995} \cite{Jaeger_2001} \cite{Minale2014a}\cite{Minale2014b}, depending upon the microstructure of the interfaces
 both between internal and ambient fluid and between gel and ambient
 fluid. Here we consider the simplest situation, which occurs if the
 gel phase is wetted by a layer of internal fluid, thus avoiding any
 interface with the ambient fluid. Otherwise, the stresses at the
 interface would have to be divided into viscous stresses and solid
 stresses. Furthermore, if a large fraction of the interface is solid,
 the situation may be complicated by the appearance of velocity slips.
 For the simple fluid-fluid interface considered here, we can assume
 that there is no velocity slip, i.e. the tangential velocity
 component is continuous across the interface. In combination with the
 condition of immiscibility, this leads to the continuity of $\mb{v}$
 across $\partial V$,
\begin{equation}
\label{eq:continuity_v}
	\mb{v}^+(\mathbf{r})=\mb{v}(\mb{r})\quad \text{for}\;  \mb{r}\in\partial V.
	\end{equation}     
A second boundary condition at $\partial V$
simply states that the interface is force free: 
\begin{equation}
	\pmb{\sigma}^+\cdot\mb{e}_r-\pmb{\sigma}^-\cdot\mb{e}_r=
2\gamma_0\mb{e}_r.%-\mb{t}^{act}\quad \text{on}\; \partial V.
	\label{eq:continuity_trac}
	\end{equation}
The balance of forces includes
the viscous tractions $\mb{t}^{\pm}=\pm\bs{\sigma}^{\pm}\cdot \mb{e}_r$ with 
$\sigma_{ij}^{-}=\sigma_{ij}(\mb{v})=-p\delta_{ij}+\eta(\partial_jv_i+\partial_iv_j)$ and $\sigma^+_{ij}=\sigma_{ij}(\mb{v}^+)$ and the Laplace pressure $2\gamma_0\mb{e}_r$ resulting from a homogeneous surface tension $\gamma_0$.

We are interested in an autonomous swimmer and hence require that it
be force and torque free, so that $\int_V \mb{f}^{act}\, d^3r =0$ and $\int_V \mb{r}\times\mb{f}^{act}\, d^3r =0$ hold.
Integrating the Brinkman equation over the volume (directly and after multiplying by $\mb{r}\times$) relates $\mb{v}_g$ and $\boldsymbol{\omega}_g$ to the internal flow field:
\begin{eqnarray}
\label{vCM}
	\int_V d^3x\;\mb{v}&=&	\int_V d^3x\;\mb{v}_g=\frac{4\pi}{3}\mb{v}_g\\
\label{wCM}
	\int_V d^3x\;\mb{r}\times\mb{v}&=&	\int_V d^3x\;
\mb{r}\times(\boldsymbol {\omega}_g\times \mb{r})= \frac{8\pi}{15}\boldsymbol{\omega}_g.
\end{eqnarray}

Brinkman's equation allows for the same expansion in vector spherical
harmonics as Stokes equation, suggesting a representation of the
active force densities in the same set of functions. The
self-propelling properties of arbitrary force densities and tractions
are entirely determined by their $l=1$ component, so that we restrict
the expansion to $l=1$:
\begin{equation}
\label{eq:active_forces}
\mb{f}^{act}=\alpha(r)Y_{10}\mb{e}_r+\beta(r)\mb{\nabla}_sY_{10}(+\gamma(r)\mb{e}_r\times\mb{\nabla}_sY_{10}.
\end{equation}
 We further
simplify the discussion by considering only $m=0$, which leads to
parallel directions of linear and angular velocity. The generalization to $m=\pm1$ and to $l\geq 1$ is straightforward.

The functions $\alpha(r), \beta(r), \gamma(r)$ %and the constants $\epsilon, \zeta, \rho$
are restricted by the requirements of vanishing total force and torque, %Eqs. (\ref{eq:forcefree}, \ref{eq:torquefree})
implying
\begin{equation}
\int_0^1 (\alpha(r)+2\beta(r))r^2\, dr=0%+\epsilon+2\zeta=0.
\label{eq:coeffforcefree}
\end{equation}
and 
\begin{equation}
\int_0^1 \gamma(r)r^3\, dr =0.
\label{eq:coefftorquefree}
\end{equation}

\section{Linear motion}
\label{sec:linear}

 As for Stokes flow, we can decompose 
active forces and the resulting flow into a chiral part ($\propto \gamma(r)$), giving rise to
rotations of the droplet around the polar axis $\mb{e}_z$ of the spherical coordinates,  and a non-chiral part, giving rise to linear motion in $\mb{e}_z$ direction.  
In terms of vector spherical harmonics  $\mb{e}_z=(\sqrt{4\pi/3})(\mb{e}_rY_{10}+\nabla_sY_{10})$, so that $\mb{v}_g=v_g\mb{e}_z=v_{g,0}\mb{e}_rY_{10}+v_{g,1}\nabla_sY_{10}$ with $v_{g,0}=v_{g,1}=\sqrt{4\pi/3}\,v_g$ (and analogously for $\boldsymbol{\omega}_g$). Note that $v_g$ and $\omega_g$ still have to be determined as part of the solution.  We first consider linear motion and hence put
$\gamma(r)=0$ %and $\rho=0$
; subsequently we are going
to discuss the rotational motion with non-vanishing $\gamma(r)$ %and $\rho$
.

\subsection{Solution of Brinkman's equation}

We use the following ansatz for the solution of Brinkman's equation:
$\mb{v}-\mb{v}_{g}=\mb{v}^{hom}+\mb{v}^{inh}$. The first term on the right hand side, $\mb{v}^{hom}$,
is the
solution of the homogeneous equation
\begin{equation}
	\nabla^2 \mathbf{v}^{hom}(\mb{r})-\frac{1}{\kappa(\phi)}\mb{v}^{hom}=0.
\label{eq:homogeneous_Brinkman}
\end{equation} 
The second term, $\mb{v}^{inh}$, denotes a special solution of the
inhomogeneous equation.
Given the restriction to  the $l=1$, $m=0$ components of the active forces, 
we represent $ \mb{v}^{hom}$ as:
\begin{equation}
  \mb{v}^{hom}=v^{hom}_0(r)Y_{10}(\theta, \varphi)\mb{e}_r+v^{hom}_1(r)\mb{\nabla}_sY_{10}(\theta, \varphi).
\end{equation}
The requirement of incompressibility, which reads explicitly $v_1(r)=(r/2)dv_0(r)/dr+v_0(r)$, can be used to reduce Eq. (\ref{eq:homogeneous_Brinkman})
to a single equation for $v^{hom}_0(r)$:
\begin{equation}\label{eq:vhom}
\left(\frac{d^2}{dr^2}+\frac{4}{r}\frac{d}{dr}- 
\frac{1}{\kappa}\right )v^{hom}_0(r)=0.
\end{equation}
This equation has two fundamental solutions, one regular at the origin
and one regular for $r\to\infty$. For the interior of the droplet we
pick the one which is regular at the origin, which is
$v^{hom}_0(r)=Au_0(r/\sqrt{\kappa})$ with
\begin{equation}
u_0(x)=\frac{1}{x}(\frac{\sinh(x)}{x^2}-
\frac{\cosh(x)}{x})= x^{-3/2}I_{3/2}(x)
\end{equation}
Here $I_{3/2}(x)$ denotes a modified Bessel function $I_\nu$ as
defined in \cite{Abramowitz_1965}.

To solve the inhomogeneous equation, we first determine the pressure, using incompressibility to write $\nabla^2p= \mb{\nabla}\cdot \mb {f}^{act}$.
The general solution (regular at the origin) is given by $p(\mb{r})= (p_0r+p^{inh}(r))Y_{10}(\theta, \varphi)$, where $p^{inh}(r)$ has to be determined from
\begin{equation}\label{eq:pinh}
\left(\frac{d^2}{dr^2}+\frac{2}{r}\frac{d}{dr}- 
\frac{2}{r^2}\right )p^{inh}=\frac{d\alpha}{dr}+\frac{2}{r}(\alpha-\beta).
\end{equation}
The inhomogeneous velocity, $\mb{v}^{inh}$ is expanded in analogy to
the homogeneous component
\begin{equation}
  \mb{v}^{inh}=v^{inh}_0(r)Y_{10}(\theta, \varphi)\mb{e}_r+v^{inh}_1(r)\mb{\nabla}_sY_{10}(\theta, \varphi).
\end{equation}
and $v^{inh}_0(r)$ is a special solution of the inhomogeneous equation
\begin{equation}\label{eq:vinh}
\left(\frac{d^2}{dr^2}+\frac{4}{r}\frac{d}{dr}- 
\frac{1}{\kappa}\right )v^{inh}_0(r)=\frac{1}{\eta}(p_0+\frac{d p^{inh}}{dr}-\alpha(r)).
\end{equation}
We will consider  explicit forms of $\alpha(r)$ and $\beta(r)$ later to illustrate the solution. 
Note that $\mb{v}^{inh}$ depends upon the pressure coefficient $p_0$, which has to be determined from boundary conditions. 
In the following we  make this dependence explicit and split  $v^{inh}_0(r)=w_0(r)-(\kappa/\eta)p_0$. 

The external flow field $v^+_0(r)=a^+/r+b^+/r^3$ is the well known solution of the Stokes problem.
Our solution thus contains four constants, $A, p_0, a^+, b^+$ and the
boundary conditions, Eq.(\ref{eq:continuity_v}) and
Eq.(\ref{eq:continuity_trac}), provide a system of four linear
equations to determine them (see Appendix~\ref{app:boundcond} for further
details). The force free condition implies the
absence of a Stokeslet, i.e. $a^+=0$.

As shown in Appendix~\ref{app:boundcond}, linear motion leads to the radial flow field $\mb{v}_r=\sqrt{3/4\pi}\,v_0(r)\cos\theta\,\mb{e}_r$ with
\begin{equation}
v_0(r)=A[u_0(r/\sqrt{\kappa})-u_0(1/\sqrt{\kappa})]+[w_0(r)-w_0(1)]+v_{g,0}.
\label{eq:v0result}
\end{equation}
The tangential flow field $\mb{v}_t=-\sqrt{3/4\pi}\,v_1(r)\sin\theta\,\mb{e}_\theta$ is determined from the incompressibility condition.  The center of mass velocity turns out to be $v_{g,0}=-(Au'_0(1)+w'_0(1))/3$  and for the constant  $A$ we find 
\begin{equation}
A= -\left(\frac{2(\eta^+-\eta)w'_0+w_0\eta/\kappa+p^{inh}}{2(\eta^+-\eta)u'_0+u_0\eta/\kappa}\right)_{r=1}.
\label{eq:A}
\end{equation}
Here $p^{inh}(r)$ and $w_0(r)$ depend on the active driving forces
which have yet to be specified.  We first discuss a monotonic force
density and subsequently introduce a nonmonotonic force, varying on a
lengthscale smaller than the size of the droplet.

\subsection{Examples of  simple force densities}

We find analytical solutions for any polynomial force density, as
discussed in detail in Appendix~\ref{app:inhomsolution}. A particularly simple example is 
$\alpha(r)=-2\beta(r)=\alpha\phi r^2$, 
which gives rise to the following inhomogeneous solution:
\begin{eqnarray}
\label{eq:pinh}
p^{inh}(r) & = & \frac{\alpha\phi}{2}r^3\\
\label{eq:w0example}
w_0(r) & = & -\frac{\alpha\phi\kappa}{2\eta}(r^2+10 \kappa) .
\end{eqnarray}
\begin{figure}[h]
	\begin{center}
\includegraphics[width=0.6\linewidth]{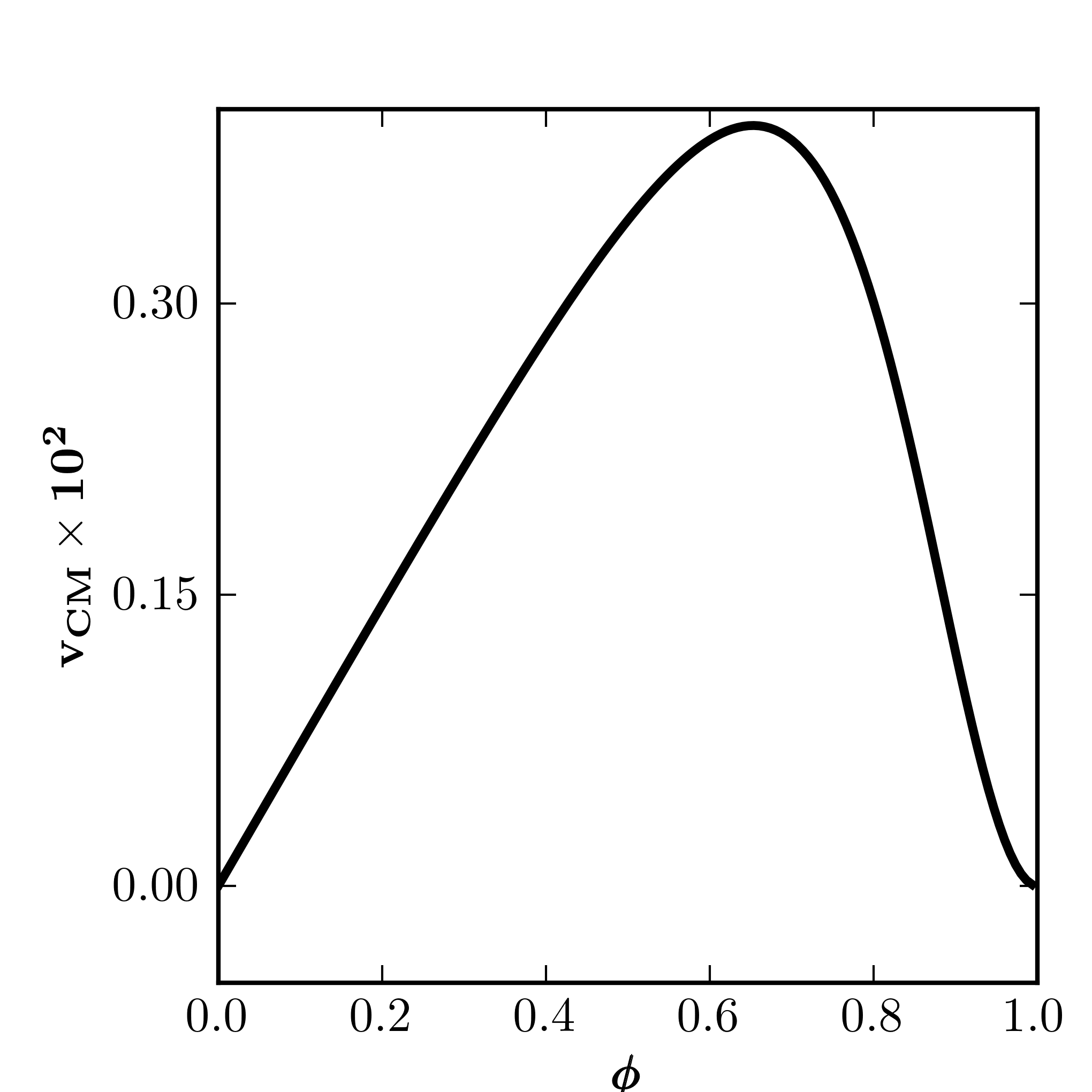}
\end{center}
\caption{\label{fig:vcm}  Center of mass velocity of the droplet vs. gel fraction $\phi$ for a force density $\alpha(r)=-2\beta(r)= \phi r^2$ and Carman-Kozeny length $\ell_0=1$.}    
\end{figure}
The resulting center of mass velocity is shown in Fig.\ref{fig:vcm} as a
function of gel fraction. The velocity vanishes for $\phi\to 0$, because the active forcing is assumed to be associated with the gel fraction and
in our simple model is taken to be proportional to $\phi$. It also
vanishes for $\phi\to 1$, when the droplet is completely rigid and
active forces cannot generate flow. These two limits imply the
nonmonotonic dependence of $v_g$ on gel fraction, which is a general feature of our model. In all examples discussed below, we put the viscosity $\eta=1$ and choose it to be ten times the viscosity of the ambient fluid.

\begin{figure}[h]
	\begin{center}
\includegraphics[width=0.45\linewidth]{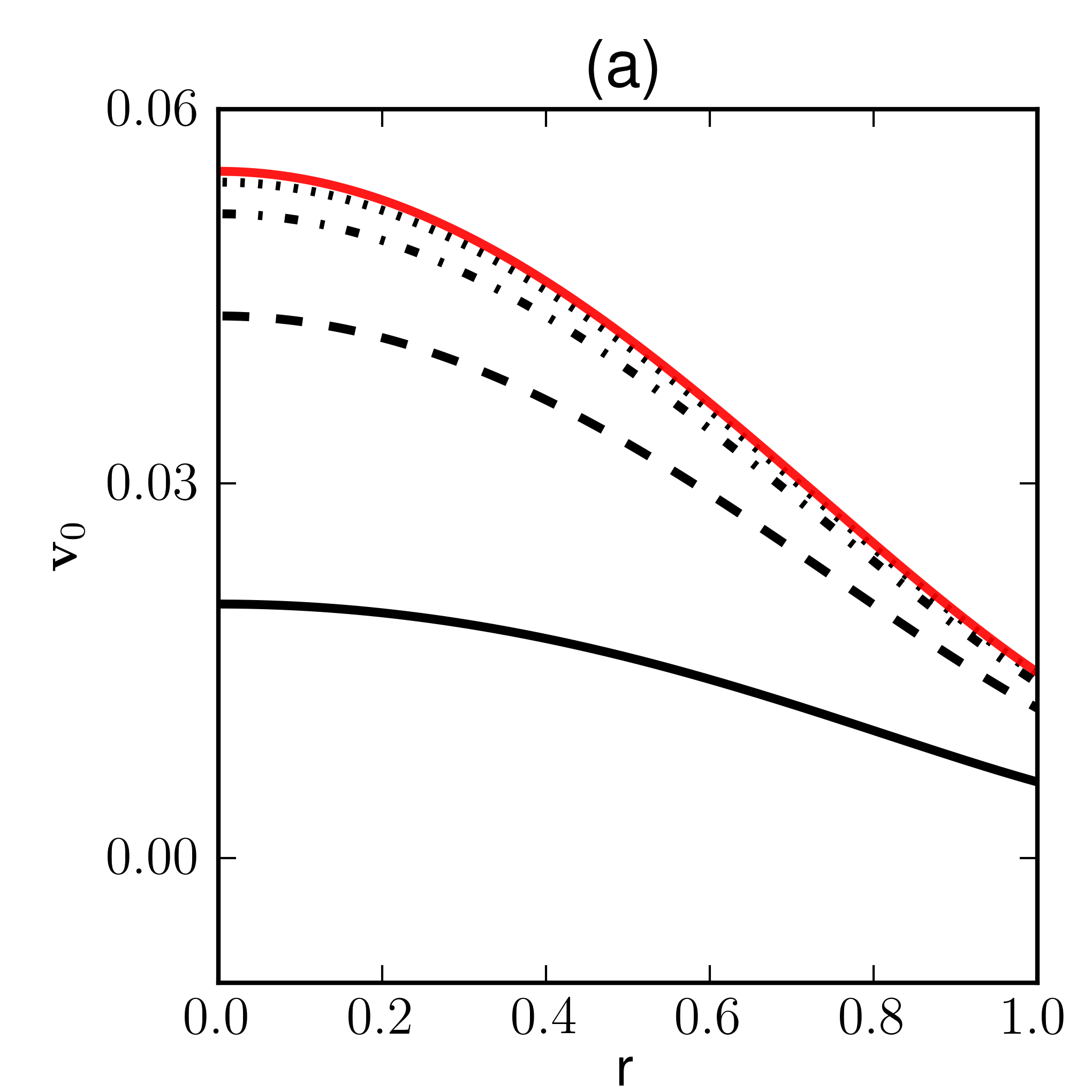}
\includegraphics[width=0.45\linewidth]{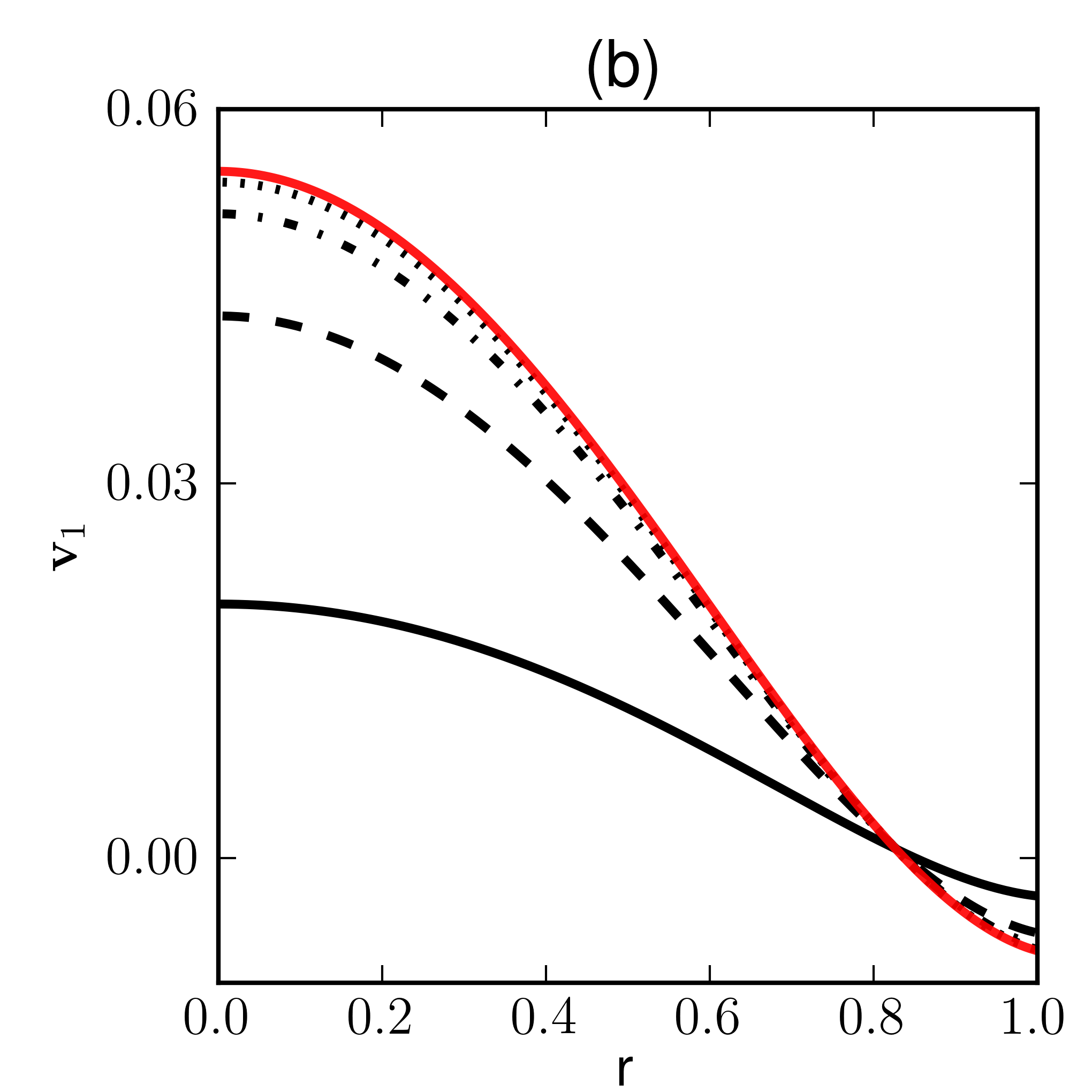}
\includegraphics[width=0.45\linewidth]{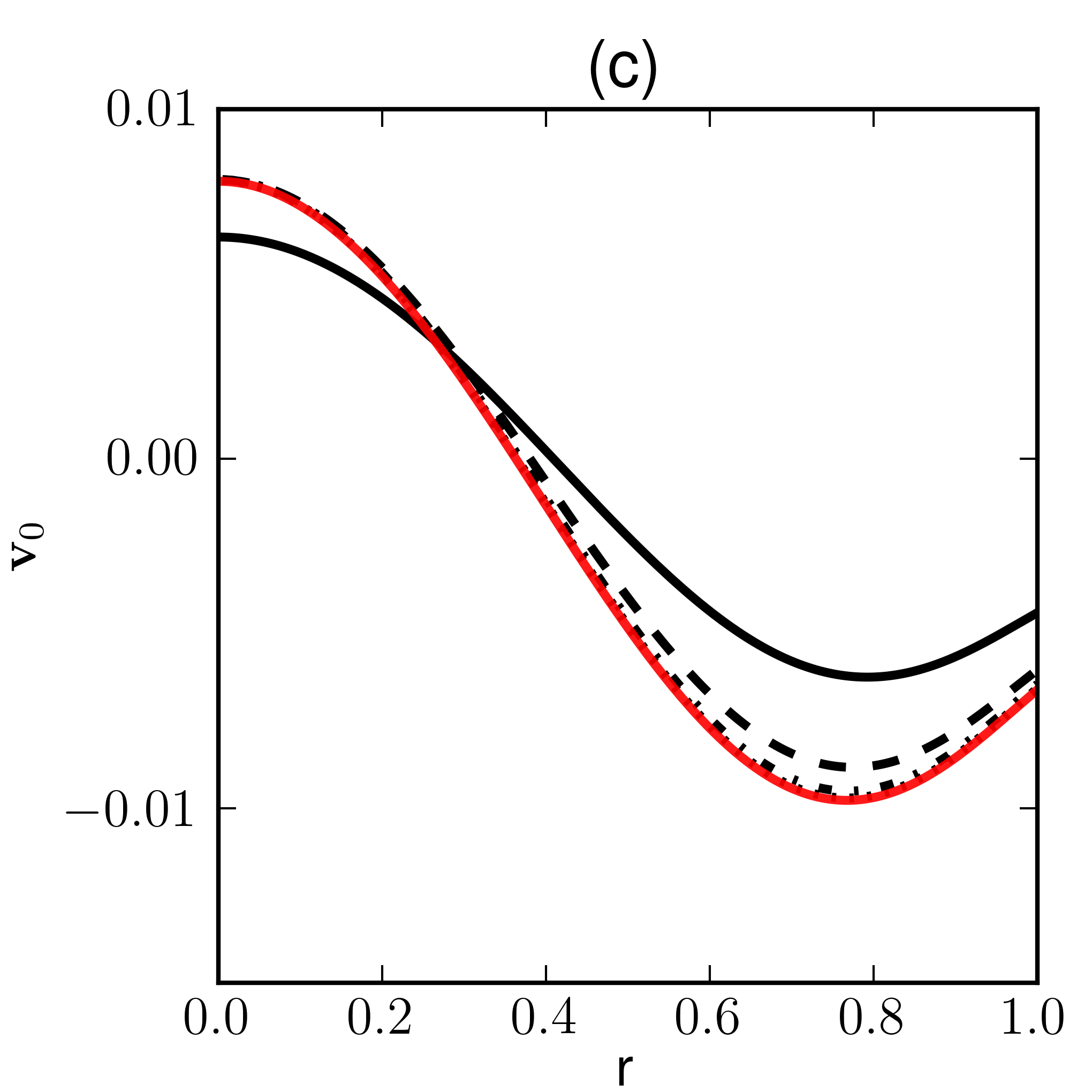}
\includegraphics[width=0.45\linewidth]{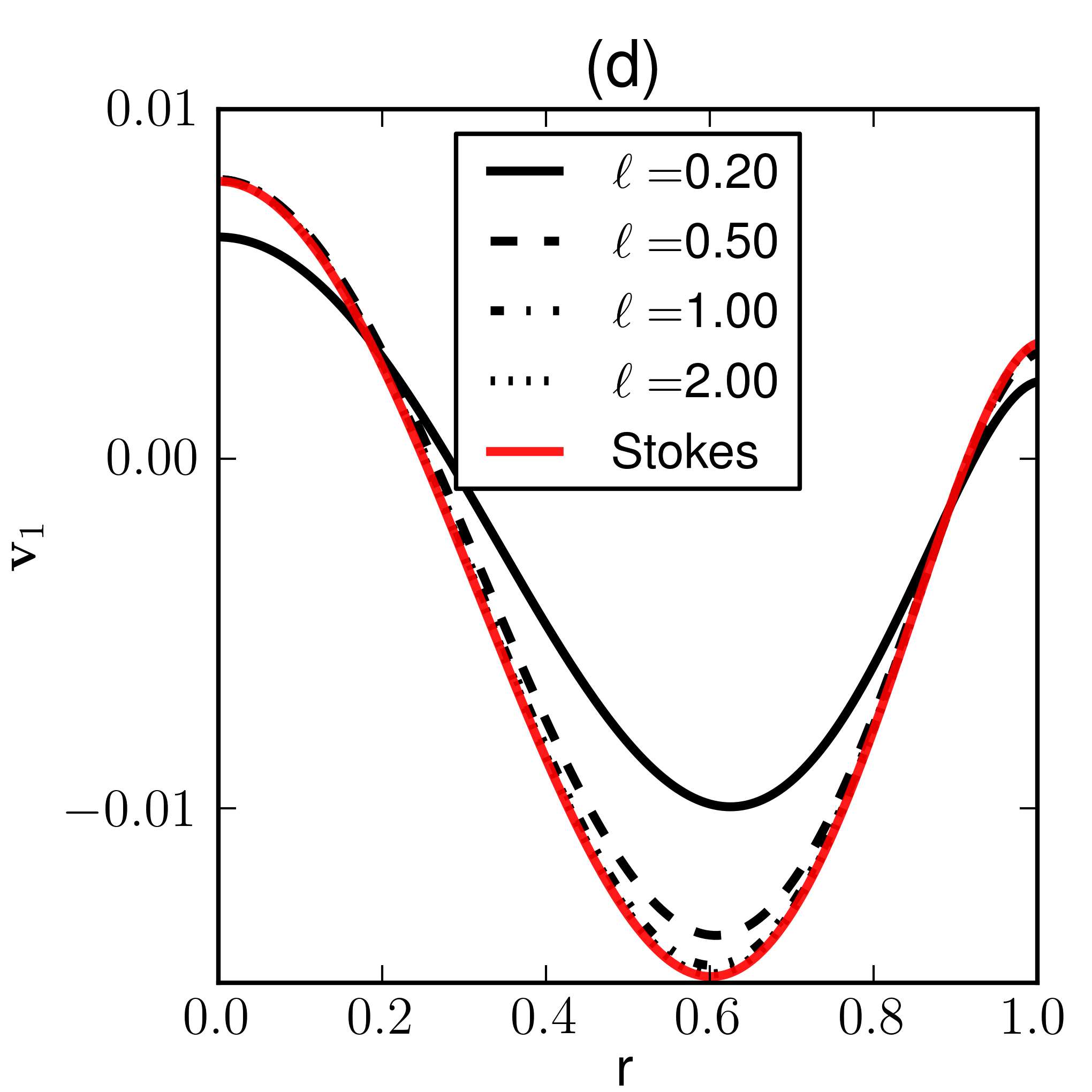}
\end{center}
\caption{\label{fig:voandv1stokes}(a) and (b): Radial velocity $v_0(r)$ and tangential velocity $v_1(r)$ for a simple force density with $\alpha(r)=-2\beta(r)=2\phi r^2$.   ((c) and (d)): for a force density  corresponding to Eq.(\ref{eq:nonmonotonic}) with $\alpha_2=10\,\phi, \alpha_4=-14\,\phi, \beta_2=-5\,\phi$ and $\beta_4=7\,\phi$. In both cases, $\phi=1/2$, so that $\ell$ and the Karman-Cozeny length $\ell_0$ coincide. The label "Stokes'' refers to $\ell\to \infty$.}
\end{figure}

Note that for $\ell\to\infty$ (\textit{Stokes limit}) and $\phi < 1$
the model must and in fact does reproduce the results for a droplet
without interaction with a rigid gel \cite{Kree_2017}.  The limit is
not immediately obvious and requires an expansion of $u_0(1/\ell)$ and
$u'_0(1/\ell)$ up to order $1/\ell^4$ to see that apparently diverging
terms cancel out.  The flow field of Stokes equation is an algebraic
function of $r$ and hence has no inherent length scale. The crossover
from Brinkman flow to Stokes flow is thus determined by the only
available lengthscale, namely the radius of the droplet, which has
been set equal to one, defining the unit of length. Hence we expect to
see Stokes flow for $\ell\gg 1$. In Fig.~\ref{fig:voandv1stokes}, we
show the flow fields, $v_0(r)$ (a) and $v_1(r)$ (b), for the simple forcing
$\alpha(r)=-2\beta(r)= \phi r^2$. The Stokes flow is seen to be a good
approximation to the Brinkman flow even for $\ell \gtrsim 1$.

A simple way to introduce a forcing lengthscale is to consider fourth order polynomials of the form, 
\begin{eqnarray}
\label{eq:nonmonotonic}
\alpha(r) & = & \alpha_2r^2+\alpha_4r^4\\
\beta(r) & = & \beta_2r^2+\beta_4r^4.
\end{eqnarray}
The parameters $\alpha_i, \beta_i, \, i=1,2$ are restricted by the
condition of Eq.(\ref{eq:coeffforcefree}), which takes on the form
$7(\alpha_2+2\beta_2)+5(\alpha_4+2\beta_4)=0$. 
The resulting inhomogeneous pressure and velocity fields are given by (see Appendix~\ref{app:inhomsolution} for details): 
\begin{eqnarray}
\label{eq:piexample}
p^{inh}(r) & = & \frac{2\alpha_2-\beta_2}{5}r^3 + \frac{3\alpha_4-\beta_4}{14}r^5\\\nonumber
\label{eq:w0example2}
w_0(r) & =  & A_4r^4+(A_2+28\kappa A_4) r^2 + (10\kappa A_2+280\kappa^2 A_4)\nonumber
\end{eqnarray}
The constants are given by $A_2=\frac{-\kappa(\alpha_2-3\beta_2)}{5\eta}$
and $A_4=\frac{-\kappa(\alpha_4-5\beta_4)}{14\eta}$
. 
In Fig.~\ref{fig:voandv1stokes} we show $v_0(r)$ (c) and $v_1(r)$ (d) for several values of $\ell$ and a forcing  corresponding to Eq.(\ref{eq:nonmonotonic}) with $\alpha_2=10\,\phi, \alpha_4=-14\,\phi, \beta_2=-5\,\phi$ and $\beta_4=7\,\phi$. The forcing is maximum for $r\approx 0.6$ and changes sign around $r\approx 0.8$,  giving rise to nonmonotonic flow for $r<1$.

\subsection{Dissipated Power}

\begin{figure}[h]
	\begin{center}
\includegraphics[width=0.6\linewidth]{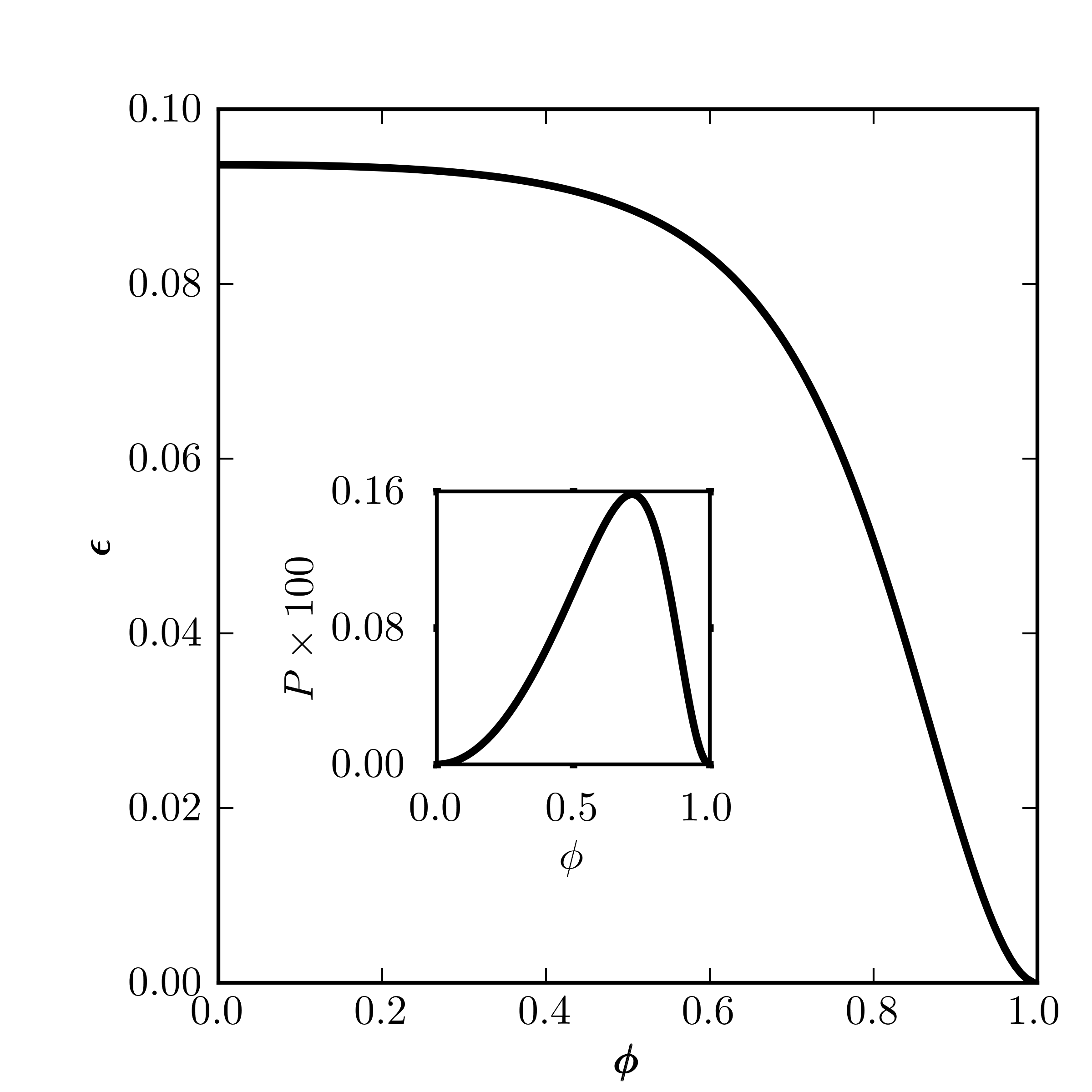}
\end{center}
\caption{\label{fig:lighthill} Lighthill efficiency vs. gel fraction for an active force density with $\alpha(r)=-2\beta(r)=\phi r^2$. Inset: Dissipated power}    
\end{figure}

In the stationary state the dissipated energy is balanced by the power
input due to the active forces. In addition to the usual viscous energy
dissipation
$\partial E_{\eta}/\partial t =\int d^3x \sigma_{i,k}\partial_k v_i$ in both,
the inside and outside fluid, we identify an additional energy
dissipation due to friction between gel and fluid: 
\begin{equation}
\frac{\partial E_{\kappa}}{\partial t} =
\frac{\eta}{\kappa}\int_{V}d^3x (\mb{v}-\mb{v}_g)^2.
\end{equation}
All 3 contributions are compensated by the power input, $P_{trans}$,
due to the  active forces 
\begin{equation}
\label{eq:power}
P_{act}=\int \mb{v}\cdot f^{act}\, dV= \int_0^1[\alpha(r)v_0(r)+2\beta(r)v_1(r)] r^2 \,  dr .
\end{equation}
%Noch mal faktoren checken!!
The power input is plotted in the inset of
Fig.~\ref{fig:lighthill} as a function of gel fraction. Not
surprisingly, do we observe the same nonmonotonic dependence as for the
flow field.

To quantify the efficiency of the propulsion mechanism we use
Lighthill's measure \cite{Lighthill_1952,Wilczek_1989},
\begin{equation}
\epsilon=\frac{P_{ext}(\mb{v}_{g})}{P_{act}},
\end{equation}
which compares the dissipated power to the corresponding power,
$P_{ext}$, needed to move a passive droplet by a constant external
force, $\mb{F}_{stall}$, with the same velocity
$\mb{v}_g$. Alternatively one can think of $\mb{F}_{stall}$ as the
force which is necessary to keep an active droplet at rest; we
therefore call it stall force.  The stall force is related to $\mb{v}_g$ via the mobility $\mu$, $\mb{F}_{stall}=\mu^{-1} \mb{v}_g$ \footnote{Note that there is a misprint in Eq. (3.10) of reference \cite{Kree_2017}, where $\mu\mb{v}_{cm}$ must be replaced by $\mu^{-1}\mb{v}_{cm}$.}, so that
$P_{ext}(\mb{v}_{g})=\mb{v}_{g}\cdot \mb{F}_{stall}=\mu^{-1}\mb{v}_g^2$.  The
mobility $\mu$ smoothly interpolates between the result for a droplet as $\phi\to 0$ and the result for a solid particle for $\phi\to 1$, as is calculated in Appendix~\ref{Stallforce}. The
efficiency is monotonically decreasing in gel fraction $\phi$ as shown
in Fig.~\ref{fig:lighthill}.

\section{Rotational motion}
\label{sec:rotation}
Let us next consider rotational motion of the droplet generated by a chiral force density 
\begin{equation}
\mb{f}^{act}_{rot}(\mb{r})=\gamma(r)\mb{e}_r\times\nabla_sY_{10}
\end{equation}  
with a single $l=1, m=0$ mode. 
The internal flow field
\begin{equation}
\mb{v}(\mb{r})=v_2(r)\mb{e}_r\times\nabla_sY_{10}
\end{equation}
is divergence free and is calculated from
\begin{equation}
\label{eq:v2}
\frac{d^2v_2}{dr^2}+\frac{2}{r}\frac{dv_2}{dr}-\frac{2}{r^2}v_2-\frac{1}{\kappa}v_2=-\frac{\gamma(r)}{\eta}+\frac{\omega_{g0}r}{\kappa} .
\end{equation}
The differential equations for $v_0$ and $v_2$ remain uncoupled, should both active forces $\mb{f}_{trans}^{act}$ and $\mb{f}^{act}_{rot}$ be present simultaneously. 

The solution procedure is completely analogous to the case of linear motion. We split $\mb{v}-\boldsymbol{\omega_g}\times\mb{r}=
(v_2+\omega_{g0}r)\mb{e}_r\times\nabla_sY_{10}=(v_2^{hom}+w_2(r))\mb{e}_r\times\nabla_sY_{10}$. In this case, the solution of the homogeneous equation is given by $
v_2^{hom}(r)=Bu_2(r/\sqrt{\kappa})$
with
\begin{equation}
\label{eq:u2}
u_2(x)=x^{-1/2}I_{3/2}(x),
\end{equation}  
and $w_2$ is a special solution of Eq.(\ref{eq:v2}) with $\omega_{g0}=0$.
The outer solution is given by $v_2^+=c^+/r^2$ and it is easily
checked, that it carries an angular momentum current, which does not
vanish at infinity. Therefore $c^+$ has to be zero for any autonomous
swimmer. This implies that the swimming (l=1) mode, which causes the
rotation of the droplet, leaves no trace in the ambient fluid.
The boundary conditions (see Appendix~\ref{app:boundcond}) thus give us two linear equations to determine  $B$ 

\begin{equation}\label{eq:B}
B=-\left(\frac{w_2-w'_2}{u_2-u'_2}\right)_{r=1}
\end{equation}
and $\omega_{g,0}$
\begin{equation}\label{eq:omega}
\omega_{g,0}=\left(\frac{w'_2u_2-w_2u'_2}{u_2-u'_2}\right)_{r=1}.
\end{equation}

A torque free force distribution cannot be achieved with a single power law, but requires a nonmonotonic force.
As for the case of linear motion we can find analytical solutions for all polynomial $\gamma(r)$ given in Appendix~\ref{app:inhomsolution}. As a simple illustrating example of a torque free force distribution we choose
\begin{equation}
\label{eq:gamma}
\gamma(r)=\gamma_0\phi\left(\frac{r}{7}-\frac{r^3}{5}\right),
\end{equation}
which leads to
\begin{equation}
v_2(r)=Bu_2(r/\sqrt{\kappa}) + \frac{\gamma_0\phi\kappa}{\eta}\left[\left(\frac{1}{7}-2\kappa\right)r-\frac{1}{5}r^3\right] -\omega_{g,0}r
\end{equation}
\begin{figure}[h]
\includegraphics[width=0.45\linewidth]{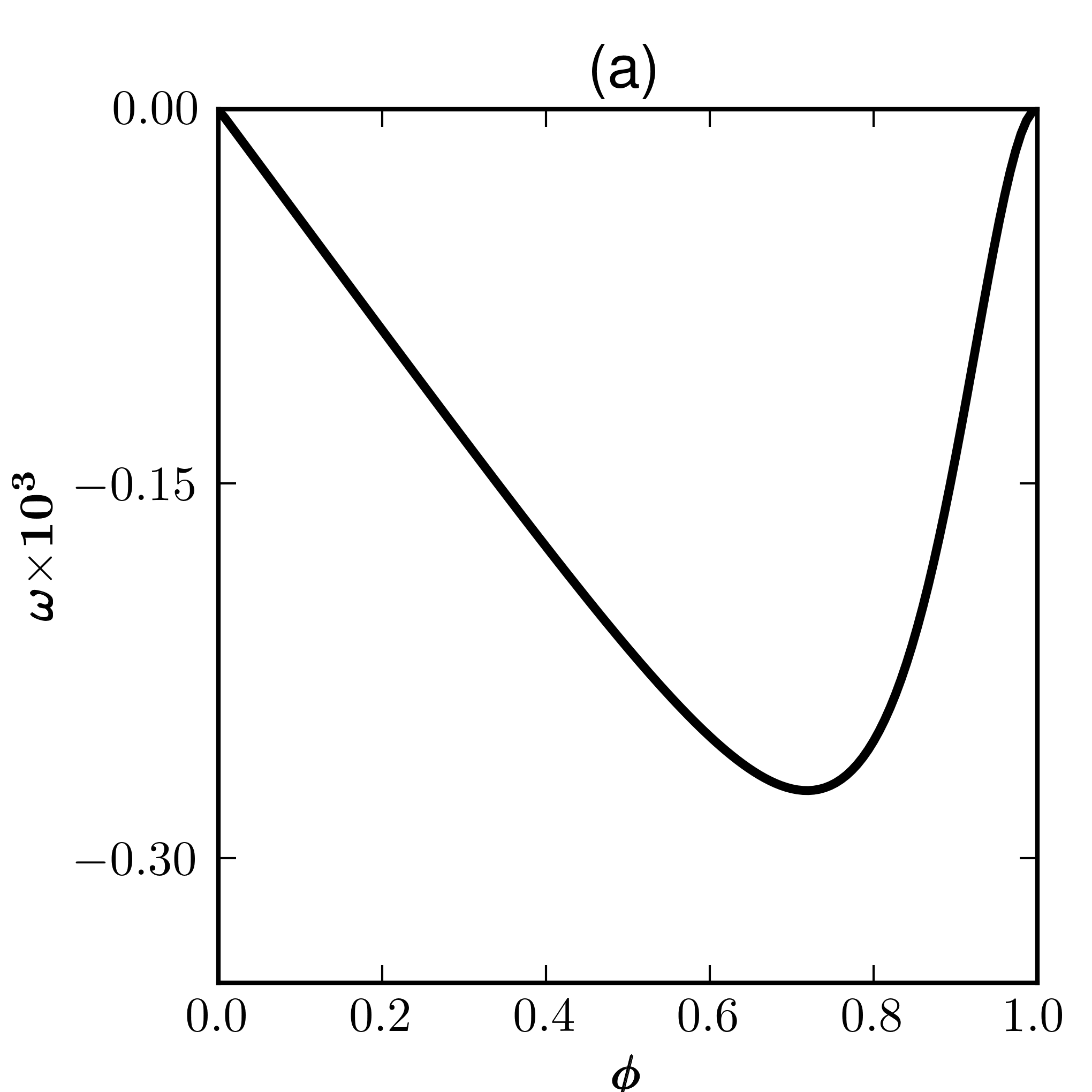}
\includegraphics[width=0.45\linewidth]{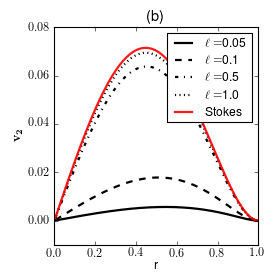}
\caption{\label{fig:omega1}(a): angular velocity versus gel fraction $\phi$ with $\ell_0=1$; (b) $v_2(r)$ versus $r$ for several $\ell$ and $\phi=1/2$; for both (a) and (b) the force density is characterised by $\gamma(r)=(5r-7r^3)2\phi$. }
%\label{fig:omega1}
\end{figure}
The angular velocity of the droplet is shown in the left part of
Fig.~\ref{fig:omega1} as a function of gel fraction. As discussed above,  the dependence is nonmonotonic, and $\omega_g$ vanishes for
$\phi=0$ and $\phi=1$, so that its overall $\phi-$dependence is similar to
the linear velocity
The tangential flow field is displayed in the right panel of
Fig.~\ref{fig:omega1}; it is nonmonotonic as a function of $r$. Its
maximum hardly depends on $\ell$ and coincides approximately with the
corresponding maximum in the forcing density.

\section{Darcy Flow versus Brinkman Flow in the limit $\ell\to 0$}
\label{sec:limits}

In the limit $\ell\to 0$ 
both, the 
velocities and the viscous stresses vanish in the interior of the droplet.
However the rescaled 
interior flow fields $\hat{\mb{v}}=\mb{v}/\ell^2$ and the droplet velocities  $\hat{v}_{g,0}=v_{g,0}/\ell^2 %\to -w'_0(1)/3\ell^2
$, 
$\hat{\omega}_{g,0}=\omega_{g,0}/\ell^2%\to\gamma(1)/\eta
$ remain finite, as will be shown explicitly below. 
 In the following, we will therefore focus on a discussion of these rescaled quantities, which we denote by an additional caret. 
 For $\ell^2=0$ the equation for $\hat{\mb{v}}$ becomes the Darcy equation
\begin{equation}
	(\hat{\mb{v}}^{D}(\mb{r})- \hat{\mb{v}}^{gel})=
\frac{1}{\eta}\Big(\mb{f}^{act}(\mb{r},\phi)-\nabla p(\mb{r})\Big),
	\label{eq:Darcy}
\end{equation} 
which is frequently used to study flows in porous media
\cite{PorousHandbook_2005}. Therefore let us consider the
approximation $\mb{v}\approx \ell^2\hat{\mb{v}}^{D}$ for $\ell\ll 1$
in detail and discuss to what extent properties of an active porous
droplet can be obtained from Darcy's equation. The second order
spatial derivative of the Brinkman equation for rescaled flow is
multiplied by $\ell^2$.  This indicates that the Darcy limit is a
singular perturbation problem, and we cannot expect that the leading
order term of Brinkman flow for $\ell\to 0$ coincides with
$\hat{\mb{v}}^D$.  In particular, the Darcy equation, when
supplemented by the incompressibility condition, is a system of first
order partial differential equations, which cannot fulfil all the
boundary conditions of the second order Brinkman equation.  Several
physical arguments have been put forward to generate a mathematically
well-posed problem for systems, which contain an interface between a
Darcy medium and a viscous fluid \cite{Ehrhardt_2012}.

First we consider the simpler case of rotational motion.
To obtain  $\hat{v}_2^{hom}$  for $\ell \ll 1$ we use the 
 asymptotic behaviour of the modified Bessel functions, $I_\nu(x)\to  \exp(x)[1-(4\nu^2-1)/8x+O(x^{-2})]/\sqrt{2\pi x}$ for $x\gg 1$. Thus the function 
 $u_2(r)$ defined in Eq.(\ref{eq:u2}) has an essential singularity
${u_2}(r/\ell)\to \ell\,\exp(r/\ell)/(\sqrt{2\pi}\,r)$  for $\ell\to 0$,
and from Eq.(\ref{eq:B}) we find that $B\to\sqrt{2\pi}\,b\,\ell^2\exp(-1/\ell)$.   Consequently, 
\begin{equation}
\hat{v}_2^{hom}(r)=\frac{Bu_2(r)}{\ell^2}\to \frac{b\ell}{r}e^{(r-1)/\ell}
\end{equation} 
and the function as well as all its derivatives vanish faster than any power of $\ell$ in the interior of the droplet, provided $x=r/\ell \gg 1$.  At the interface $r=1$, $\hat{v}_2^{\hom}(1)\to b\ell$, and it contributes to the tractions as $\hat{t}_2^{hom}(1)\to \eta \hat{v}_2^{\hom}(1)/\ell=\eta b$ -- even at $\ell=0$.  Thus,  the rescaled  traction of the homogeneous Brinkman flow  does not vanish for $\ell\to 0$ and contributes to the boundary conditions  Eqs.(\ref{eq:rotboundcond1}, \ref{eq:rotboundcond2}), which become $b\ell=\hat{\omega}_{g,0}-\hat{w}_2(1)\to 0$ and $b-\hat{\omega}_{g,0}-\hat{w}'_2(1)=0$ for $\ell \to 0$.  For Darcy flow, the homogeneous solution is strictly 0, so that $b=0$ holds from the outset. Consequently  there is no consistent solution for $\hat{\omega}^D_{g,0}$ . If we require $\hat{\omega}^D_{g,0}=\hat{\omega}_{g,0}$ the   Darcy flow  fulfils the boundary condition for the rescaled velocity $\hat{v}_2$ (see Eq.(\ref{eq:rotboundcond1})), but there remains a slip in the tangential traction $\hat{t}_2$ at the interface, i.e. Eq.(\ref{eq:rotboundcond2}) is violated.

\begin{figure}[h]
\includegraphics[width=0.45\linewidth]{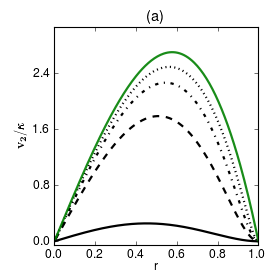}
\includegraphics[width=0.45\linewidth]{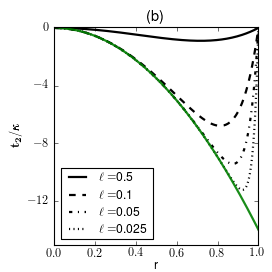}\\
\caption{\label{fig:v2andt1Darcy} $v_2/\kappa$ and $t_2/\kappa$ vs. $r$ for different values of $\ell=\sqrt{\kappa}$  in comparison to the Darcy solution, demonstrating the uniform convergence of the Brinkman flow towards the Darcy flow (a) and the existence of a boundary layer in the tractions (b).}
\end{figure}

In Fig.~\ref{fig:v2andt1Darcy} we show  $\hat{v}_2(r)$ and   $\hat{t}_2(r)$ for
the special forcing $\gamma(r)$ of Eq.(\ref{eq:gamma}) in comparison to the flow field, $\hat{v}_2^{D}(r)=-\hat{\omega}^{D}_{g,0}r+\gamma(r)/\eta$, of the Darcy equation.  Note that we have chosen the value $\hat{\omega}^D_{g,0}=\gamma(1)/\eta$ of the Brinkman flow in the Darcy flow to enable a comparison.  As expected from our discussion, the convergence of the Brinkman flow to the Darcy flow is uniform in $r$, but the rescaled tractions develop a boundary layer.
Thus our discussion has revealed that for small $\ell$ the Darcy solution is a good approximation to the Brinkman flow in a frame of reference co-rotating with the gel,  it also approximates the corresponding viscous traction in the interior outside of a boundary layer at the interface,  but it cannot explain the self-propelling property.

The case of linear motion is similar, but slightly more complicated
due to the pressure contribution. As for the rotational motion, we can
use the asymptotics of the Bessel function to obatin the homogeneous
solution of the Brinkman equation, $v_0^{hom}=Au_0(r/\ell)$, in the
limit $\ell\to 0$. As for the rotational motion 
${u_0}(r/\ell) \to \ell^2/(\sqrt{2\pi}\, r^2)\,\exp(r/\ell),$
has an essential singularity at $\ell=0$. With  $A=\sqrt{2\pi}a\ell^2\exp(-1/\ell)$, we find in the limit $\ell\to 0$:
\begin{equation}
\hat{v}_0^{hom}(r)=\frac{Au_0(r)}{\ell^2} \to \frac{a\ell^2}{r^2}\,e^{(r-1)/\ell}.
\end{equation}
Analogous to the case of  rotational motion,  $v_0^{hom}$ 
generates a finite rescaled tangential traction  $\hat{t}^{hom}_1=t_1^{hom}(1)/\ell^2\to \eta (v_0^{hom})''/(2\ell^2)\to\eta  a/2$ at the interface in the limit $\ell\to 0$. 
The contribution to the radial traction becomes
$t_0^{hom}(1)/\ell^2\to -(p_0+p^{inh}(1))/\ell^2+2\eta \hat{w}'_0(1)$.
From the solution for polynomial force densities (see
Appendix~\ref{app:inhomsolution}) we find that
$p_0+p^{inh}(1) = O(\ell^2)$, and from Eq. (\ref{eq:linsys1}) one
reads off that $p_0\to \eta(a\ell^2+w_0(1))$. Thus, even in the limit
$\ell\to 0$ the rescaled pressure on the interface contains a
non-vanishing term proportional to $a$, which results from the
homogeneous flow. If we put $a=0$, as required for the Darcy equation,
no consistent solution for $\hat{v}_{g,0}=b^+/\ell^2$ can be obtained
from the boundary conditions Eqs.(\ref{eq:linsys1}-\ref{eq:linsys3}).

Let us illustrate
this in more detail for the simple force density
$\alpha(r)=-2\beta(r)=\alpha\phi r^2$.  From Eqs.(\ref{eq:pinh},
\ref{eq:w0example}) one sees that $p(r)=p_0r+(\alpha\phi/2)r^3$ and
$\hat{w}_0(r)=-\alpha\phi(r^2+10\ell^2)/(2\eta)$. 
On the other hand, solving the Darcy equation for this particular force density we find $p^D(r)=p_0^Dr+(\alpha\phi/2)r^3$
and $\hat{w}_0^D(r)=-\alpha\phi r^2/(2\eta)$. The boundary condition (\ref{eq:linsys1}) determines $p_0\to \eta (a\ell^2+\hat{w}_0(1))
=\eta a\ell^2-\alpha\phi(1+10\ell^2)/2$ and $p_0^D=-\alpha\phi/2$.  Hence the difference in pressure $\Delta p(r)=p(r)-p^D(r)$ is solely due to the homogeneous pressure and explicitly given by $\Delta p(r)=\ell^2(\eta a-5\alpha\phi)+O(\ell^{4})$.
It dominates the difference between the radial tractions of Brinkman and Darcy flow 
\begin{equation}
\hat{t}_0-\hat{t}_0^{D}\to \frac{(p_0^{D}-p_0)r}{\ell^2}=(\eta a-5\alpha)r,
\end{equation}
and is \emph{not} restricted to a boundary layer but a bulk effect. 
Inserting these asymptotic
forms for $\ell\ll 1$ into the boundary conditions
Eqs.(\ref{eq:linsys2}, \ref{eq:linsys3}) we get
$ \hat{v}_{g,0} = \alpha\phi/3\eta + O(\ell^2)$and
$\eta^+\hat{v}_{g,0} = \eta a -3\alpha\phi +O(\ell^2)$ which are
obviously inconsistent for $a=0$. 

\begin{figure}[h]
	\begin{center}
\includegraphics[width=0.45\linewidth]{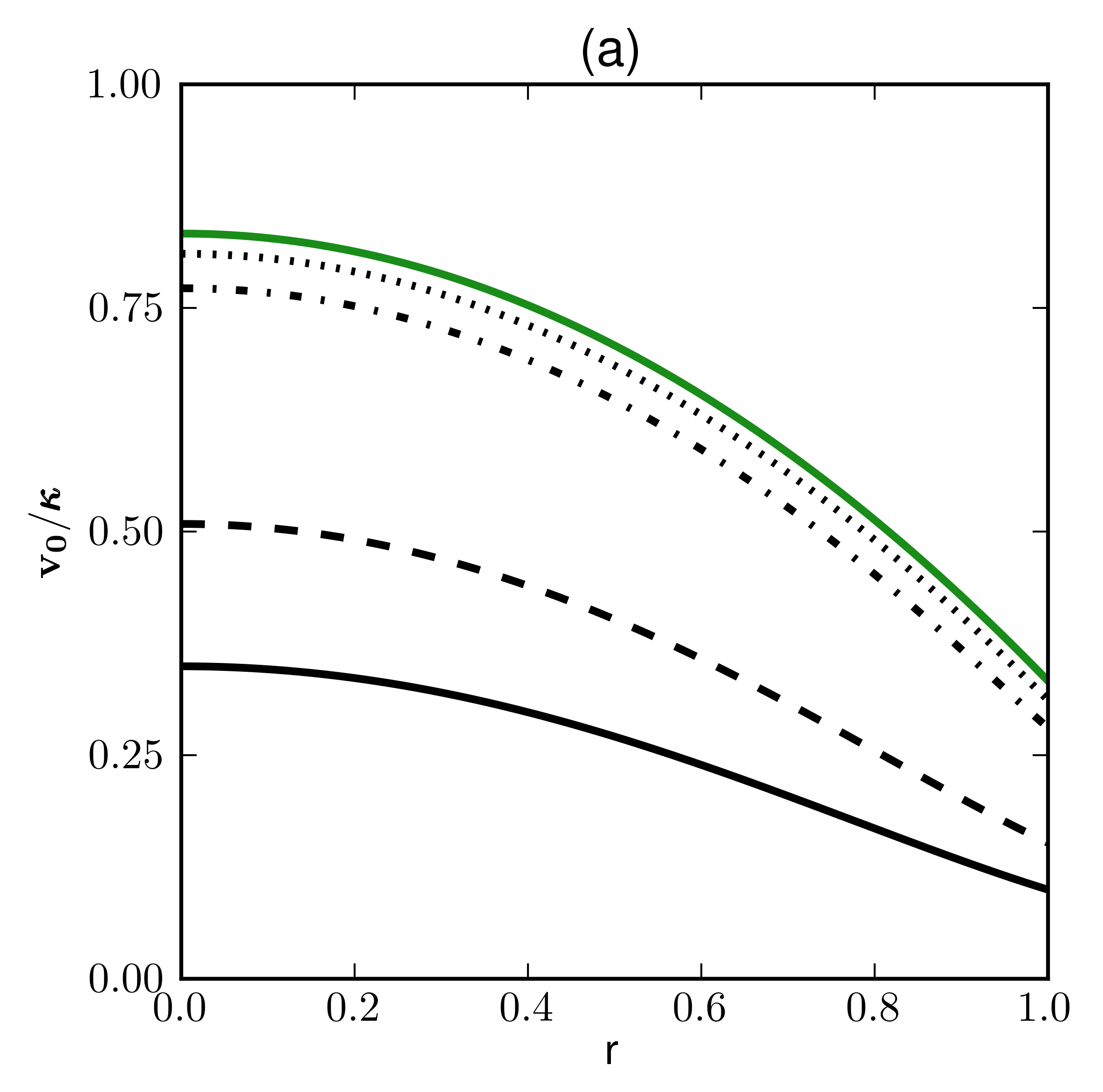}
\includegraphics[width=0.45\linewidth]{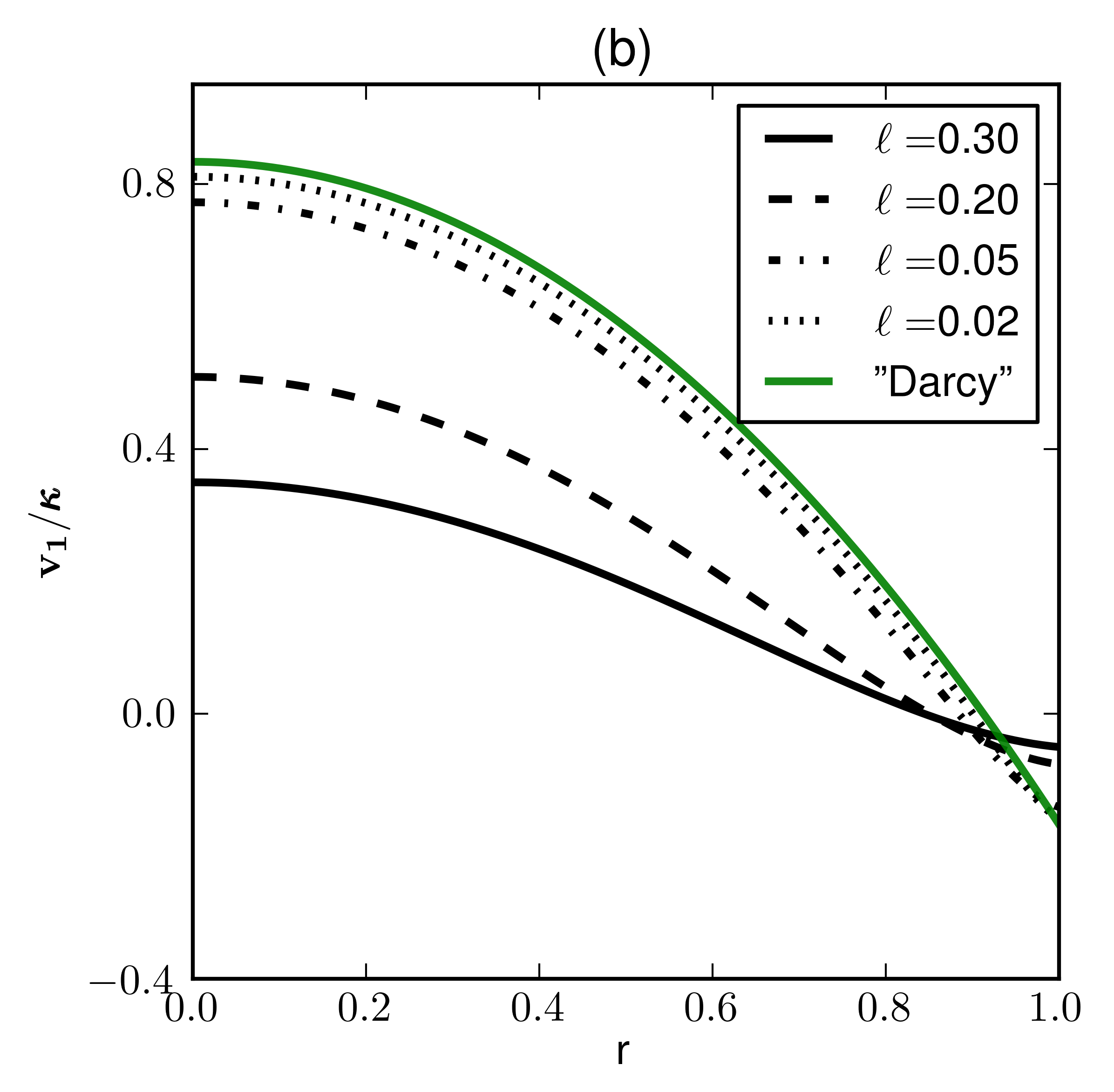}
\end{center}
\caption{\label{fig:voandv1Darcy} Radial (a) and tangential (b) flow fields, $v_0,v_1$,   in comparison to the Darcy solution, showing uniform convergence of the flow fileds as $\ell \to 0$.}
  \end{figure}

\begin{figure}
  \begin{center}
\includegraphics[width=0.45\linewidth]{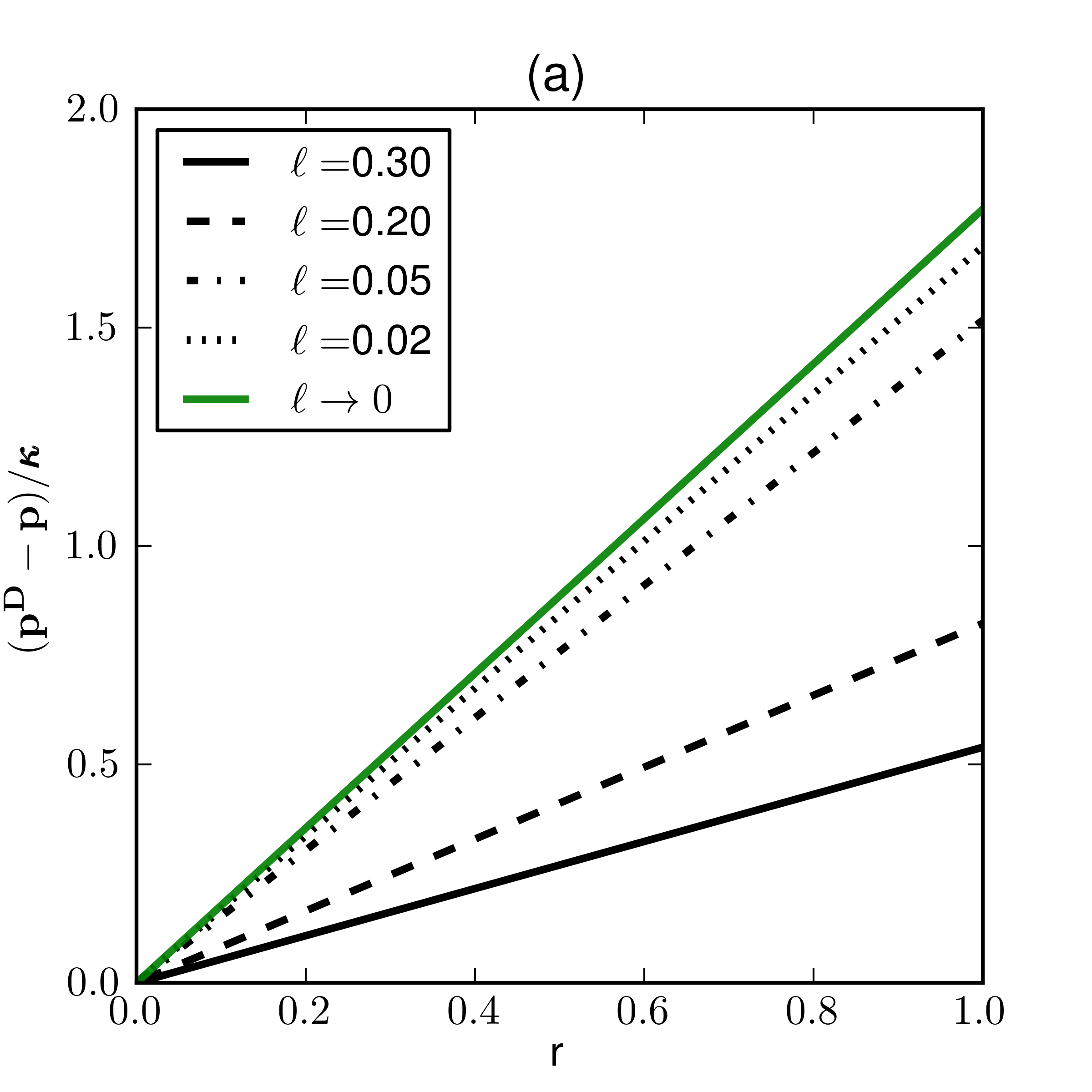}
\includegraphics[width=0.45\linewidth]{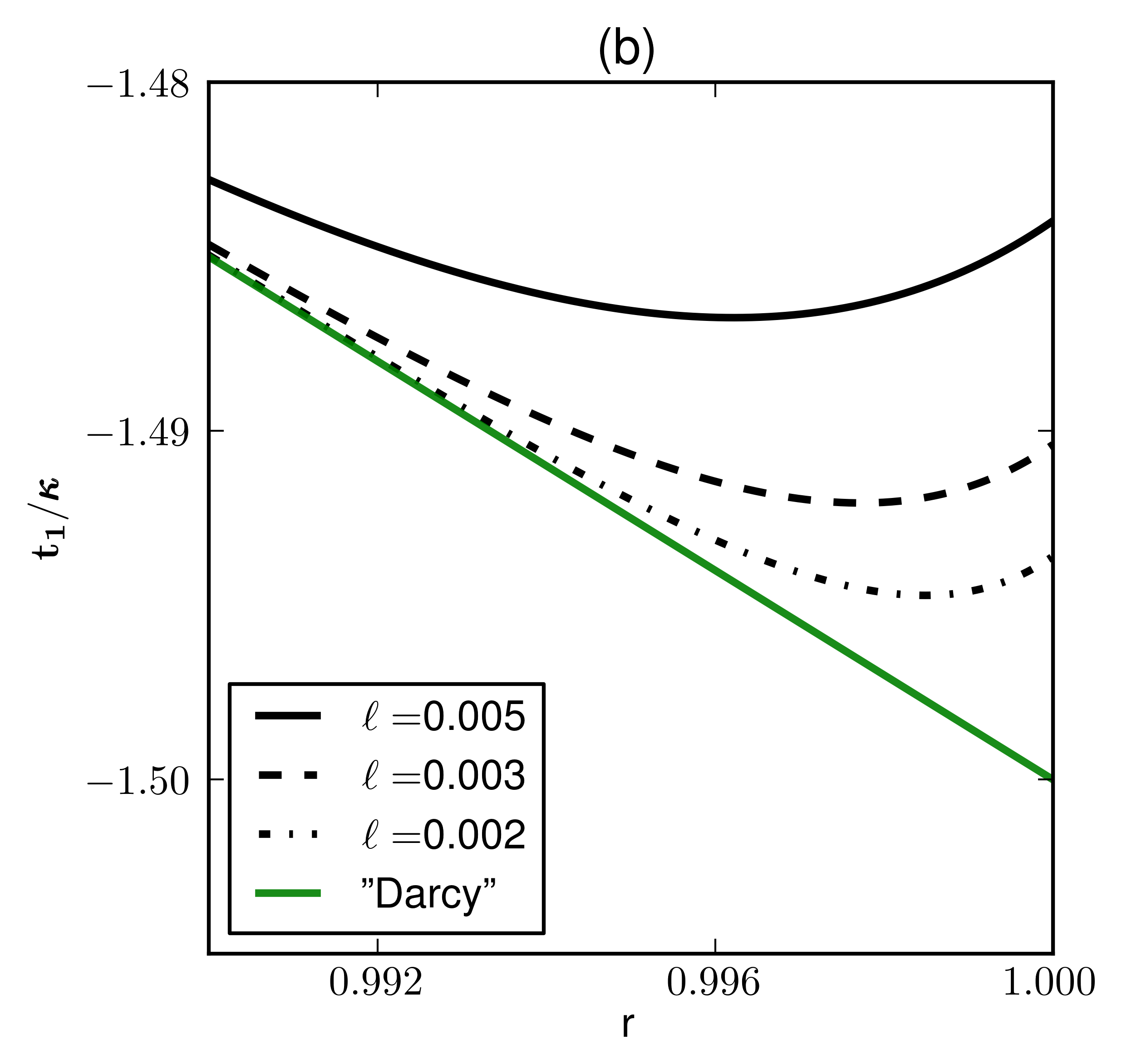}
\end{center}

\caption{\label{fig:pandt1Darcy} (a) $(p^D-p)/\kappa$ is the asymptotic difference of rescaled radial tractions, shown vs r. This difference is not confined to a boundary layer at the interface.  (b) tangential traction $t_1$ for several values of $\ell$ in comparison to the Darcy solution, displaying the emergence of a boundary layer for decreasing $\ell$.}
\end{figure}

The approach of the rescaled flow fields and tractions towards their
limits is shown in Fig.~\ref{fig:voandv1Darcy} and
Fig.~\ref{fig:pandt1Darcy}. To allow for a comparison with Darcy flow, 
$\hat{v}_0^D(r)=-(\alpha\phi r^2/2-p^D_0)/\eta+v^D_{g,0}$, we have set 
$v^D_{g,0}=v_{g,0}$.
The radial and tangential flow fields are
observed to approach the Darcy solution uniformly in the interior
(Fig.~\ref{fig:voandv1Darcy}). On the other hand the rescaled pressure
and radial tractions of the Brinkman solution differ from the Darcy
solution in all of the interior bulk, as demonstrated in the left part
of Fig.\ref{fig:pandt1Darcy}. The difference in tangential tractions is
restricted to a boundary layer as for the torsional tractions for the
rotational motion.
   

\section{Conclusions and Outlook}
\label{sec:conclusions}
We have proposed a biphasic model for an active swimmer, consisting of
a gel and a viscous fluid. Building on previous work on porous media,
we model the fluid interacting with a rigid gel by Brinkman's
equation. The permeability of the gel, $\kappa$, defines a
characteristic lengthscale $\ell=\sqrt{\kappa}$, which depends on the
gel fraction $\phi$. It decreases monotonically with gelfraction: If
the gel fills the droplet completely, the lengthscale vanishes. In the
opposite limit, for very small gel fraction the lengthscale $\ell$
diverges and we recover Stokes flow.

Activity is modelled by active forces, residing on the gel. We assume
their magnitude to be proportional to the gel fraction. The resulting
linear and rotational propulsion velocity of the swimmer are nonmonotonic
functions of the gel fraction $\phi$. Both velocities vanish for
$\phi\to 0$, because the strength of the active forces go to zero, and
they also vanish as $\phi\to 1$, because the forces cannot generate
internal flow in the space filling rigid gel. Hence there is an
optimal gelfraction, giving rise to maximal linear and rotational
propulsion. We also discuss energy dissipation, which in addition to
viscous dissipation inside and outside of the droplet has an
additional contribution from the viscous drag between gel and fluid.
Modelling active body forces allows us to introduce another
lengthscale which can give rise to nonmonotonic flow as a function of $r$ inside the droplet.

As the gel fraction approaches one, the lengthscale goes to zero and
one might naively expect that the flow follows Darcy's equation.
However this is not the case, because $\ell \to 0$ is a singular limit
of the Brinkman equation: The solution of the homogeneous equation,
$\mb{v}^{hom}(r)$, aquires an essential singularity, so that the
homogeneous flow and all its derivatives vanish faster than any power
of $\ell$ in the interior of the droplet.
% However, the homogeneous
%flow at the boundary, $\mb{v}^{hom}(r=1)\propto \ell^3$, gives rise
%to tractions $\mb{t}^{hom}(r=1)\propto \ell^2$ which contribute to
%momentum transfer across the interface. The resulting velocity of
%self-propulsion is $\mb{v}_{gel}\propto
%\ell^2$.
For Darcy`s equation the boundary conditions cannot be fulfilled and there is
no consistent solution for $\mb{v}_{gel}$. Hence the Darcy equation
cannot explain self-propulsion.

If we enforce the correct $\mb{v}_{gel}$ in the Darcy equation, then
the flow fields within the droplet are well approximated by Darcy flow
for $\ell \to 0$.  However the pressure in the bulk of the droplet is
not accounted for by Darcy`s equation and the tangential and torsional
tractions are incorrect in a boundary layer, reflecting the
incompatibility of Darcy`s equation with the boundary conditions.

Even though we have explicitly discussed body forces only, the
treatment of surface tractions is straightforward. Similarly, the
restriction to $m=0$ was chosen for simplicity and can easily be
lifted.  As far as self-propulsion is concerned the expansion in
angular momenta can be restricted to $l=1$, the only mode which
contributes to self-propulsion. A correct description of the flow
fields inside and outside of the droplet requires inclusion of all $l$,
which is easily feasable.

Our approach can be extended in several ways. Our model of the gel as
a completely rigid porous structure, should be improved. A first step
is to consider an elastic gel, a next step could be a viscoelastic
medium including dynamics of the gel. So far we have only discussed
the simplest forcing. However, the approach is general enough to allow
for a more realistic modelling of biological forcing mechanisms, such
as an ensemble of motors, moving along a filamentary structure. Work
along these lines is in progress.

%%
%% For one-column wide figures use
%\begin{figure}
%% Use the relevant command for your figure-insertion program
%% to insert the figure file.
%% For example, with the option graphics use
%\resizebox{0.75\textwidth}{!}{%
%  \includegraphics{leer.eps}
%}
%% If not, use
%%\vspace{5cm}       % Give the correct figure height in cm
%\end{figure}
%%
%% For two-column wide figures use
%\begin{figure*}
%% Use the relevant command for your figure-insertion program
%% to insert the figure file. See example above.
%% If not, use
%\vspace*{5cm}       % Give the correct figure height in cm
%\end{figure*}
%\begin{table}
%\begin{tabular}{lll}
%\hline\noalign{\smallskip}
%\end{tabular}
%% Or use
%\vspace*{5cm}  % with the correct table height
%\end{table}
%%
%% The section below may be edited at your convenience to acknowledge 
%% each author's contribution to the manuscript.
%% You may remove it if you are a single author.
%%
\section{Authors contributions}
All the authors were involved in the preparation of the manuscript.
All the authors have read and approved the final manuscript.
%
% BibTeX users please use
% \bibliographystyle{}
% \bibliography{}
%
% Non-

\appendix
\numberwithin{equation}{section}
\section{Boundary conditions}
\label{app:boundcond}

The interface $\partial V$ of the spherical droplet at $r=1$ is a free
boundary, and the flow field is continuous across $\partial V$.  The
velocity components 
inside the droplet are given by
$v_a(r)=v^{hom}_a(r)+ v^{inh}_a(r)+v_{g,0}$ (for $a=0,1$), 
 $v_0=Au_0(r/\sqrt{\kappa})
+w_0(r)-(\kappa/\eta)p_0+v_{g,0}$ and
$v_2(r)=Bu_2(r/\sqrt{\kappa})+w_2(r)-\omega_{g,0}r$.  The
incompressibility condition determines $v_1$ in terms of
$v_0$: $v_1(r)=v_0(r)+r v'_0(r)/2$.  Here and in the following the
prime denotes the derivative with respect to $r$.  The external flow
field is determined by $v^+_0(r)=a^+/r+b^+/r^3, v_2=c^+/r^3$ and the
linear velocity $v_{g,0}$ coincides with $v^+_0(r=1)$.

The viscous tractions  $\mb{t}^{\pm}=\pm\bs{\sigma}^{\pm}\cdot \mb{e}_r$
in a spherical geometry are represented in the from \cite{Kree_2017}:
\begin{equation}
\mb{t}=t_0(r)Y_{10}\mb{e}_r+t_1(r)\mb{\nabla}_sY_{10}+t_2(r)\,\mb{e}_r\times\mb{\nabla}_sY_{10}.
\end{equation}
The flow inside the droplet gives the following contributions
\begin{eqnarray}
\label{eq:t0}
t_0^-(r) & = & -p_0(r)-p^{inh}(r)+2\eta v'_0(r)\\
\label{eq:t1}
t_1^-(r) & = & \eta\Big(v'_1+(v_0(r)-v_1(r))/r\Big)\\
\label{eq:t2}
t_2^-(r) & = & \eta(v'_2-v_2/r).
\end{eqnarray}
whereas the outside flow and hence the tractions $\mb{t}^{+}$ are the
same as in \cite{Kree_2017}.

  Using these quantities, the boundary
conditions take on the form of 6 linear equations for the unknowns
$a^+, b^+$ $c^+, A, p_0, B$. The two equations involving $v_2$ and
$t_2$ are decoupled from the rest, which takes on the form:

\begin{eqnarray}
 \label{eq:bc1}
0 & = & Au_0+w_0-\frac{\kappa}{\eta}p_0\\
\label{eq:bc2}
-\frac{a^++3b^+}{2} & = & A\left(u_0+\frac{u'_0}{2}\right)\\\nonumber 
+ & & \left(w_0+\frac{w'_0}{2}\right) -\frac{\kappa}{\eta}p_0\\
\label{eq:bc3}
-3\eta^+(a^++2b^+) & = & -p_0-p^{inh}+2\eta(Au'_0+w'_0)\\
\label{eq:bc4}
3\eta^+b^+ & = & \eta(v'_1-v_1+v_0).
\end{eqnarray}
The remaining equations become
\begin{eqnarray}
\label{eq:bc5}
c^+ & = & Bu_2+w_2-\omega_{g,0}\\
\label{eq:bc6}
-3\eta^+c^+ & = & \eta[B(u'_2-u_2)+w'_2-w_2]%+\rho
\end{eqnarray}
Here all functions are taken at $r=1$ and  the prime denotes the derivative with respect to $r$, for example $u'_0=du_0/dr=(1/\sqrt{\kappa})(du_0(x)/dx)_{x=1/\sqrt{\kappa}}$. 

From the force-free and torque-free condition, we find that $a^+=0$
and $c^+=0$.  The vanishing of $c^+$ is obvious from the $r^{-3}$
dependence of the outer flow field, which leads to $r^{-4}$ dependence
of the viscous stress and thus to a non-vanishing total torque acting
on the entire system, in contradiction to the torque-free
condition. The vanishing of $a^+$ can be inferred by adding up two
times Eq. (\ref{eq:bc4}) and Eq.(\ref{eq:bc3}), which gives
\begin{equation}
\label{eq:avanish}
-3\eta^+a^+=t_0^-+2t_1^-
\end{equation}
On the r.h.s. of this equation we can use
\begin{eqnarray}
\int_{\partial V} \mb{t}\, d^2S & = & \int_{\partial V} (t_0 Y_{10}\mb{e}_r+t_1\nabla_sY_{10})\, d^2S\\\nonumber
&=& \int_V \nabla\cdot \boldsymbol{\sigma}\, d^3V=\sqrt{\frac{4\pi}{3}}(t_0+2t_1)\mb{e}_2.
\end{eqnarray}
Inserting $\nabla\cdot\boldsymbol{\sigma}=-\mb{f}$, one can replace $t_0+2t_1$ by $-\int_0^1(\alpha+2\beta)r^2\, dr$, and thus the r.h.s of Eq.(\ref{eq:avanish}) becomes the force-free condition Eq. (\ref{eq:coeffforcefree}).

Using $a^+=0$ and subtracting Eq.(\ref{eq:bc1}) from Eq. (\ref{eq:bc2} ) 
leaves us with three linearly
independent equations, 
 which can be written in the form
\begin{eqnarray}
\label{eq:linsys1}
p_0 & = & \frac{\eta}{\kappa}(Au_0+w_0)\\
\label{eq:linsys2}
b^+ =v_{g0} & = & -\frac{1}{3} (u'_0A +w'_0)\\
\label{eq:linsys3}
6\eta^+b^++2\eta u'_0 A - p_0 & =&-2\eta w'_0+p^{inh}
\end{eqnarray}
Solving the last two equations for $A$, yields the result given in
Eq.~\ref{eq:A} in the main text.  Note that with Eq.(\ref{eq:bc1}) the
radial velocity field $v_0(r)$ becomes
\begin{eqnarray}
v_0(r) & = &Au_0(r)+w_0(r)+v_{g,0}-\frac{\kappa p_0}{\eta}\\\nonumber &=&A(u_0(r)-u_0(1))\\\nonumber 
& + &(w_0(r)-w_0(1))+v_{g,0}.
\end{eqnarray}
In this form it is given in the main text in Eq.(\ref{eq:v0result}).

For rotational motion Eqs. (\ref{eq:bc5}- \ref{eq:bc6}) simplify for $c^+=0$:
\begin{eqnarray}
\label{eq:rotboundcond1}
(Bu_2 -\omega_{g,0}+w_2)_{r=1} & = & 0\\
\label{eq:rotboundcond2}
(Bu'_2  -\omega_{g,0} + w'_2)_{r=1} & = & 0.
\end{eqnarray}
with the explicit solution given in Eqs.~(\ref{eq:B},\ref{eq:omega}).

\section{Flow fields from polynomial force densities}
\label{app:inhomsolution}
Here we discuss special solutions of Eqs.~(\ref{eq:pinh}, \ref{eq:vinh})  and Eq.(\ref{eq:v2}), for force densities with polynomial $\alpha(r), \beta(r), \gamma(r)$, like  $\alpha(r)=\alpha_0+\alpha_1r+\alpha_2 r^2 +\cdots$.

The pressure equation (\ref{eq:pinh}) has polynomial solutions, provided we put $\alpha_0=\beta_0$. (For $\alpha_0\neq\beta_0$, terms $\propto r\ln r$ will also occur.)  For  nth order monomials, $\alpha_nr^n$ and $\beta_nr^n$, a particular solution, regular at $r=0$,  is given by 
\begin{equation}
p^{inh}(r)=\frac{(n+2)\alpha_n-2\beta_n}{n(n+3)}r^{n+1}=p_n^{inh}r^{n+1}.
\end{equation}
Note that for $n=2$ and $n=4$ this produces Eq.(\ref{eq:piexample}).

All inhomogeneities in Eq.(\ref{eq:vinh}) are thus polynomials  $\sum_{n=0} D_{n}r^{n}$ with  $D_{n}=(p_n^{inh}(n+1)-\alpha_n)/\eta$.  For example, $D_2=(\alpha_2-3\beta_2)/5\eta$ and $D_4= (\alpha_4-5\beta_4)/14\eta$.

For every monomial of even order, $D_{2m}r^{2m}$, it is easy to find a solution from the ansatz $$w_0=A_{2m}r^{2m}+A_{2m-2}r^{2m-2}+\cdots A_2r^2- A_0.$$ It  leads to 
$A_{2k-2}=2k(2k+3)\kappa A_{2k}$  for $k=1,2\cdots m$, which allows to determine the $A_{2k}$ starting from $A_{2m}=-\kappa D_{2m}$. 

So, for example, for force densities $\propto r^2$, corresponding to  $m=1$, this leads to 
\begin{equation}
w^{m=1}_0=A_2(r^2+10\kappa),
\end{equation}
with $A_2=-\kappa D_2$,
and for $m=2$ we get
\begin{equation}
w^{m=2}=A_4(r^4+28\kappa r^2 + 280\kappa^2).
\end{equation}
In the main text, we considered  both $r^2$ and $r^4$ terms to be present, so that 
\begin{eqnarray}\label{eq:w2andw4}
  w_0 & = & w_{04}r^4 + w_{02}r^2+w_{00}\\\nonumber
  & = & A_4r^4+(A_2+28\kappa A_4) r^2 + 
(10\kappa A_2+280\kappa^2 A_4)
\end{eqnarray}

For odd order monomials $D_{2m+1}r^{2m+1}$ a polynomial ansatz for $w_0$ is not sufficient. Extending the ansatz by terms $A_{-1}r^{-1}$ and  $A_{-3}r^{-3}$ does lead to  solutions, but they are not regular at $r=0$.  To construct solutions, which stay regular in the droplet's interior, one has to add a homogeneous solution, which compensates the singular terms.  A regular solution takes on the form 
\begin{eqnarray}
w_0& = &C(r/\sqrt{\kappa})^{-3/2}K_{3/2}(r/\sqrt{\kappa}) +(-8r^{-3}\\\nonumber
& + &4r^{-1} +  r)A_1 + \cdots -\kappa D_{2m+1}r^{2m+1}.
\end{eqnarray}
 and the coefficients $A_{2k+1}$ can be determined in analogy to the case of even order inhomogeneities. 
 
Starting from Eq. (\ref{eq:v2}) and polynomials $\gamma(r)$ we can proceed analogously to construct $w_2$. Odd order monomials $\gamma_{2n+1}r^{2n+1}$ lead to polynomial  solutions $w_2=B_{2n+1}r{2n+1}+B_{2n-1}r^{2n-1}+\cdots B_1 r$ with $B_{2n+1}=\kappa\gamma_{2n+1}/\eta$  and $B_{2k-1}=\kappa[(2k+1)(2k+2)-2]B_{2k+1}$. For even order monomials, the Polynomial Ansatz has to be extended  by terms $B_0+B_{-2}r^{-2}$ and a regular solution has to be constructed by adding a homogeneous solution  $F(r/\sqrt{\kappa})^{-1/2}K_{3/2}(r/\sqrt{\kappa})$.

\section{Stall force and mobility}
\label{Stallforce}

%We only consider $m=0$, so that ${\mb v}_g$ points in the
%z-direction. 
To move a passive droplet with  velocity $v_g$ in z-direction, one applies a
constant force density $\mb{f}_{stall}=f\mb{e}_z=-\nabla U$, which gives rise to a total force $F=4\pi f/3$.

The inhomogeneity
in Eq.~\ref{eq:Brinkman1} is then given by
$\nabla p(\mb{r})- \mb{f}_{stall}=\nabla \tilde{p}$, so that the stall
force can be absorbed into a modified pressure $\tilde{p}=p-U$. The
latter fulfils Laplace equation, whose solution is given by
$\tilde{p}(\mb{r})=\tilde{p}_0rY_{10}(\theta, \varphi)$ with
$\tilde{p}_0=(p_0-\sqrt{4\pi/3}f)$.  The only inhomogeneity in
Eq.~\ref{eq:vinh} is $\tilde{p}_0$, so that $w_0=0$. 
The set of Eqs.~\ref{eq:bc1} to \ref{eq:bc4} then reads
\begin{eqnarray}
0 & = & Au_0-\frac{\kappa}{\eta}\tilde{p}_0\\
-\frac{a^++3b^+}{2} & = & A\left(u_0+\frac{u'_0}{2}\right)
 -\frac{\kappa}{\eta}\tilde{p}_0\\
-3\eta^+(a^++2b^+) & = & -p_0+2\eta Au'_0\\
3\eta^+b^+ & = & A\eta(u_0'+u_0''/2).
\end{eqnarray}
and has to be solved for the four unknowns $A,p_0,a^+,b^+$. Due to the presence of a force, there is a Stokeslet with $a^+=\sqrt{4\pi/3}f/(3\eta^+)$. The center of mass velocity is given by $v_g=\sqrt{3/(4\pi)}(a^++b^+)=\mu F$ with the mobility 
\begin{equation}
\mu=\frac{1}{4\pi\eta^+}\,\left(1+\frac{\eta(u_0/\kappa-2u_0')/3}
{2u_0'(\eta-\eta^+)-\eta u_0/\kappa}
\right).
\end{equation}
In the Stokes limit $\kappa\to \infty$ and 
$
u_0/(\kappa u'_0)\to 5.
$
This reproduces the well-known mobility $\mu=(\lambda +1)/[2\pi\eta^+(3\lambda +2)]$ of a fluid droplet with viscosity contrast $\lambda=\eta/\eta^+$.
In the Darcy limit the terms proportional to $u'_0$ can be neglected and one recovers the mobility of a solid particle. 
\vspace{2em}
\providecommand{\newblock}{}

%\bibliographystyle{iopart-num}
%\bibliography{Microswimmer}
\end{document}